\documentclass[10pt]{article}
\usepackage{fullpage,amssymb,amsmath}
\usepackage[dvips]{epsfig,graphicx}

\def\##1{{\bf #1}}
\def\=#1{\underline{\underline{#1}}}

\def\+#1{\underline{\bf #1}}
\def\*#1{\underline{\underline{\bf #1}}}

\def\_#1{\underline{#1}}

\def\r#1{(\ref{#1})}
\def\l#1{\label{#1}}
\def\c#1{\cite{#1}}

\def\le{\left(}
\def\ri{\right)}
\def\les{\left[}
\def\ris{\right]}
\def\lec{\left\{}
\def\ric{\right\}}

\def\.{\mbox{ \tiny{$^\bullet$} }}

\def\epso{\epsilon_{\scriptscriptstyle 0}}

\def\eps{\epsilon}

\def\det{\mbox{det}}
\def\adj{\mbox{adj}}
\def\Tr{\mbox{tr}}

\begin{document}

\LARGE
\begin{center}
{\bf The homogenization of orthorhombic piezoelectric composites by
the strong--property--fluctuation theory}

\vspace{10mm} \large

Andrew J. Duncan\footnote{Corresponding author.  E--mail:
Andrew.Duncan@ed.ac.uk.},  Tom G. Mackay\footnote{E--mail: T.Mackay@ed.ac.uk.}\\
\emph{School of Mathematics and
   Maxwell Institute for Mathematical Sciences\\ University of Edinburgh, Edinburgh EH9
3JZ, UK}

\vspace{3mm}

 Akhlesh  Lakhtakia\footnote{E--mail: akhlesh@psu.edu}\\
 \emph{ NanoMM~---~Nanoengineered Metamaterials Group\\ Department
  of Engineering Science and Mechanics\\
Pennsylvania State University, University Park, PA 16802--6812, USA}

\end{center}

\vspace{4mm}

\normalsize

\begin{center}
{\bf Abstract}
\end{center}

The linear strong--property--fluctuation theory (SPFT) was developed
in order to estimate the constitutive parameters of certain
homogenized composite materials (HCMs) in the long--wavelength
regime. The component materials of the HCM were generally
orthorhombic $mm2$ piezoelectric
 materials,  which were randomly
distributed as oriented ellipsoidal particles. At the second--order
level of approximation, wherein a two--point correlation function
and its associated correlation length characterize the component
material distributions, the SPFT estimates of the HCM constitutive
parameters were expressed in terms of numerically--tractable
two--dimensional integrals. Representative numerical calculations
revealed that: (i) the lowest--order SPFT estimates are
qualitatively similar to those provided by the corresponding
Mori--Tanaka homogenization formalism, but differences between the
two estimates become more pronounced as the component particles
become more eccentric in shape; and (ii) the second--order SPFT
estimate provides a significant correction to the lowest--order
estimate, which reflects dissipative losses due to scattering.

\vspace{4mm}

\normalsize


\section{Introduction}

Since piezoelectric materials can convert electrical energy to
mechanical energy, and vice versa, they  are of considerable
technological importance. However, bulk  piezoelectric materials
commonly exhibit physical properties which render them unsuitable
for particular applications. For example, certain ceramics exhibit
strong piezoelectric properties but their weight, malleability and
acoustic impedance are not suitable for transducer applications
\c{Ting}. Accordingly, composite piezoelectric materials  are often
more technologically attractive \c{Zhang}; and these can be found in
a host of applications such as
 in transducers, sensors, actuators and energy harvesting devices, for
example \c{Dam,Swallow}. Furthermore, the recent proliferation of
multifunctional metamaterials \c{Walser}~---~which often take the
form of homogenized composite materials (HCMs), exhibiting exotic
constitutive properties \c{M05}~---~presents interesting
possibilities for piezoelectric HCMs.

While the estimation of  elastodynamic or electromagnetic
constitutive parameters of HCMs is a challenging task, especially
for anisotropic HCMs, the estimation of constitutive parameters of
piezoelectric HCMs is more challenging due to  the coupling of
elastodynamic and electromagnetic fields. Numerous homogenization
formalisms  have been proposed for piezoelectric HCMs, many of which
build upon   Eshelby's landmark description of the elastodynamic
response of a single ellipsoidal particle immersed in an infinite
homogeneous medium \c{Eshelby,Giordano}. For example, the
Mori--Tanaka \c{Mori-Tanaka,DunnTaya,Huang_Kuo}, self--consistent
and differential approaches \c{DunnTaya2}~---~and combinations of
these \c{Odegard}~---~feature prominently in the literature. In the
following we present a fundamentally different approach to
estimating the constitutive properties of piezoelectric HCMs, based
on the
 strong--property--fluctuation theory (SPFT).
A key feature of the SPFT homogenization approach~---~which
distinguishes it from other more conventional approaches~---~is the
accommodation of higher--order characterizations of the
distributional statistics of the HCM's component materials.

The origins of the SPFT lie in wave propagation studies for
continuously random mediums \cite{Ryzhov}. It was later adapted to
estimate the electromagnetic \cite{TK81,ML95,spft_form}, acoustic
\cite{Zhuck_acoustics} and elastodynamic \cite{spft_zhuck1}
constitutive parameters of HCMs. Within the SPFT, the estimation the
HCMs  constitutive parameters arises by successive refinements to
the constitutive parameters of a homogeneous comparison medium.
Iterations are expressed in terms of correlation functions
describing the spatial distributions of the component materials. In
principle, correlation functions of arbitrarily high--order may be
incorporated; but, in practice, the SPFT is most often implemented
at the second--order level of approximation, wherein a two--point
correlation function and its associated correlation length
characterize the component material distributions.

We establish here the linear, second--order SPFT appropriate to
orthorhombic $mm2$ piezoelectric HCMs, arising from component
materials which are randomly distributed as oriented ellipsoidal
particles. The theoretical development builds upon the corresponding
development of the orthotropic elastodynamic SPFT
\cite{spft_zhuck1,DML}. A representative numerical example is used
to illustrate the theory, and results are compared with those from
the well--established Mori--Tanaka formalism.

\section{Theory}\label{Theory_section}
\subsection{Preliminaries}\label{Prelim}

In the following, we consider piezoelectric materials described by
constitutive relations of the form \c{Nye,Auld_book}
\begin{equation}
\left.
\begin{array}{l}
\sigma_{ab} = C_{abmn} S_{mn} - e_{nab} E_n\\
D_a = e_{amn} S_{mn} + \eps_{an} E_n
\end{array}
\right\}, \l{CRs_1}
\end{equation}
wherein the elastic strain $S_{mn}$ and the electric field $E_n$ are
taken as independent variables, which are related to the stress
$\sigma_{ab}$ and dielectric displacement $D_a$ via the elastic
stiffness tensor $C_{abmn}$ (measured in a  constant electric
field), the piezoelectric tensor $e_{nab}$ (measured at a constant
strain or electric field), and the dielectric tensor $\eps_{an}$
(measured at a constant strain). Here, and hereafter, tensors are
represented in plain font and lowercase tensor indexes range from 1
to 3, with  a repeated index implying summation.

We develop the SPFT in the frequency domain. Accordingly the
complex--valued representations of the stress, strain and
electromagnetic fields have an implicit $\exp \le - i \omega t \ri$
dependency on time $t$, with $\omega$ being the angular frequency
and $i = \sqrt{-1}$. The possibility of dissipative behaviour is
thereby accommodated via the imaginary parts of complex--valued
constitutive parameters.

The constitutive relations \r{CRs_1} are more conveniently expressed
in the symbolic  form
\begin{equation} \l{CRs_tensor}
\breve{\sigma}^{}_{aB} = \breve{C}_{aBMn}^{} \breve{S}_{Mn}^{},
\end{equation}
where the extended stress symbol
\begin{equation} \label{form_stress}
\breve{\sigma}_{aB}^{} = \left\{
\begin{array}{lcr}
  \sigma_{ab}^{}, && B=b=1,2,3 \\
  D_a^{}, && B=4 \\
\end{array}\right.,
\end{equation}
the extended stiffness symbol
\begin{equation}\label{form_stiffness}
\breve{C}_{aBMn}^{} = \left\{
\begin{array}{lcr}
  C_{abmn}^{}, && B=b=1,2,3; \: M=m=1,2,3 \\
  e_{nab}^{},  && B=b=1,2,3; \: M=4 \\
  -e_{amn}^{},  && B=4; \:M=m=1,2,3 \\
  \epsilon_{an}^{}, && B,M=4 \\
\end{array}\right.,
\end{equation}
and the extended strain symbol
\begin{equation} \label{form_strain}
\breve{S}_{Mn}^{} = \left\{
\begin{array}{lcr}
  S_{mn}^{}, && M=m=1,2,3 \\
  E_n^{}, && M=4 \\
\end{array}\right. .
\end{equation}
Here, and hereafter, uppercase  indexes range from 1 to 4. Note that
the extended quantities defined in eqs. \r{form_stress},
\r{form_stiffness} and \r{form_strain} are not tensors~---~these are
simply symbols which are introduced to allow a compact
representation of the piezoelectric constitutive relations
\cite{DunnTaya}.

In developing the SPFT appropriate to piezoelectric HCMs, it is
expedient to express the constitutive relations \r{CRs_tensor} in
matrix--vector form as
\begin{equation}
\mbox{\boldmath$\breve{\#\sigma}$}^{} =
\breve{\*C}^{}\cdot\breve{\#S}^{},
\end{equation}
wherein $\mbox{\boldmath$\breve{\#\sigma}$}$ and $\breve{\#S}$ are
column 12--vectors representing the  extended stress and extended
strain symbols, respectively, and $\breve{\*C}^{}$ is a 12$\times$12
matrix which represents the extended stiffness symbol.
 Here, and hereafter, matrixes are denoted by double underlining and
bold font, whereas vectors are in bold font with no underlining. For
use later on, we note that
 the $pq$th entry of a matrix $\*A$ is
written as $\les \, \*A \, \ris_{pq}$, while the $p$th entry of a
vector $\#b$ is written as $\les \, \#b \, \ris_p$. Accordingly, the
matrix entry $\big[ \, \*A \cdot \*B \,\big]_{pr}= \big[\, \*A \,
\big]_{pq}\big[ \, \*B \, \big]_{qr}$, the vector entry $\big[ \,
\*A \cdot \#b\big]_{p}= \big[ \, \*A \, \big]_{pq}
\big[\#b\big]_{q}$, and the scalar $\#a \cdot \#b =  \les \#a \ris_p
\les \#b \ris_p$. The adjoint, determinant, inverse, trace and
transpose of a matrix $\*A$ are denoted by $\adj \le \, \*A\, \ri$,
$\det \le \, \*A \, \ri$, $\*A^{-1}$,  $\Tr \le \, \*A \, \ri$ and
$\*A^T$, respectively. The $n\times n$ null matrix is written as
$\*0_{\,n \times n }$.

Our concern in this article is with orthorhombic $mm2$ piezoelectric
materials \c{Nye,Auld_book}. For this symmetry class, the extended
stiffness matrix has the block matrix form
\begin{equation} \l{stiffness_mat_def}
\breve{\*C} = \le \begin{array}{cc} \*C & - \*e^T \vspace{8pt} \\
\*e & \mbox{\boldmath$\*\epsilon$}
\end{array} \ri,
\end{equation}
where the 9$\times$9 stiffness matrix $\*C$ may be expressed as
\begin{equation}
\*C = \le \begin{array}{ccc} \*C_{\,a} & \*0_{\,3\times3} & \*0_{\,3\times3}  \\
\*0_{\,3\times3} & \*C_{\,b} & \*C_{\,b} \\
\*0_{\,3\times3} & \*C_{\,b} & \*C_{\,b}
\end{array}\ri,
\end{equation}
with the 3$\times$3 symmetric matrix components
\begin{equation}
\*C_{\,a} = \le \begin{array}{ccc} C_{11} & C_{12} & C_{13}
 \\
C_{12} & C_{22} & C_{23}
\\
C_{13} & C_{23} & C_{33}
\end{array}\ri, \qquad \*C_{\,b} = \le \begin{array}{ccc} C_{44} & 0 &
0
 \\
0 & C_{55} & 0
\\
0 & 0 & C_{66}
\end{array}\ri;
\end{equation}
 while the 9$\times$3
piezoelectric matrix $\*e$ may be expressed as
\begin{equation}
\*e = \le \begin{array}{ccccccccc} 0 & 0 & 0 & 0& e_{15} &0 &0  & e_{15} &0\\
0 & 0 & 0 & e_{24} & 0 & 0  & e_{24} & 0 & 0\\
e_{31} & e_{32} & e_{33} & 0 &0  & 0 &0 &0 &0
\end{array}
\ri \end{equation} and the 3$\times$3 dielectric matrix
$\mbox{\boldmath$\*\epsilon$} $ as
\begin{equation}
\mbox{\boldmath$\*\epsilon$} = \le \begin{array}{ccc} \eps_{11} &0 &
0 \\ 0 & \eps_{22} &0 \\
0 &0 & \eps_{33} \end{array} \ri.
\end{equation}
The correspondence between the tensor/extended symbol representation
and the
 matrix--vector representation  is described in Appendix~A.

In an analogous fashion,
 the material density $\rho$ may be represented via the extended
density symbol
\begin{equation}
 \breve{\rho}_{BM} = \left\{
\begin{array}{lcr}
  \rho^{},  && B=M=1,2,3 \\
  0, && \textrm{otherwise} \\
\end{array}\right.,
\end{equation}
which  has the 4$\times$4 extended matrix counterpart
$\mbox{\boldmath$\*{\breve{\rho}}$}^{}$ with entries
\begin{equation}
 \les \, \mbox{\boldmath$\*{\breve{\rho}}$}^{}
\, \ris_{MP}= \breve{\rho}_{MP}.
\end{equation}

\subsection{Component materials}

We consider the homogenization of a composite comprising two
component materials,  labelled as component material `1' and
component material `2'. In general, both components are homogeneous,
orthorhombic $mm2$ piezoelectric materials, characterized by the
stiffness tensors $C^{(\ell)}_{abmn}$, piezoelectric tensors
$e^{(\ell)}_{nab}$, dielectric tensors $\eps^{(\ell)}_{an}$ and
densities $\rho^{(\ell)}$ $(\ell = 1,2)$. In conformity with the
notational practices introduced in \S\ref{Prelim}, the component
materials are also described by the extended stiffness symbols
$\breve{C}^{(\ell)}_{aBMn}$ (and their 12$\times$12 matrix
equivalents $\breve{\*C}^{(\ell)}$) and extended density symbols
$\breve{\rho}^{(\ell)}_{BM}$  (and their 4$\times$4 matrix
equivalents $\mbox{\boldmath$\*{\breve{\rho}}$}^{(\ell)}$).

 The component materials are
randomly distributed  as identically--oriented, conformal,
ellipsoidal particles. The principal axes of the ellipsoidal
particles are aligned with the Cartesian axes. Thus, the surface of
each ellipsoidal particle may be parameterized by the vector
\begin{equation} \l{r_shape}
\#r^{(e)} = \eta \*U \cdot\hat{\#r},
\end{equation}
where $\eta$ is a linear measure of size, $\hat{\#r}$ is the radial
unit vector and the diagonal shape matrix
\begin{equation}\label{U_shape}
\*U=\frac{1}{\sqrt[3]{abc}}\left(
\begin{array}{ccc}
a & 0 & 0 \\
0 & b & 0 \\
0 & 0 & c \\
\end{array}
\right),\qquad \qquad  (a,b,c \in\mathbb{R}^{+}).
\end{equation}

Let $V$ denote the space occupied by the composite material. Then $V
=  V^{(1)} \cup V^{(2)}$ where $V^{(1)}$ and $V^{(2)}$ contain the
two component materials labelled as `1' and `2', respectively, and
$V^{(1)}  \cap V^{(2)} = \emptyset$. The distributional statistics
of the component materials are described in terms of moments of the
characteristic functions
\begin{equation}
\Phi^{(\ell)} (\#r) = \left\{ \begin{array}{ll} 1, & \qquad \#r \in
V^{(\ell)},\\ & \qquad \qquad \qquad \qquad \qquad \qquad (\ell=1,2) . \\
 0, & \qquad \#r \not\in V^{(\ell)}, \end{array} \right.
\end{equation}
 The first statistical moment of
$\Phi^{(\ell)}$, i.e.,
\begin{equation} \l{vf}
 \langle \, \Phi^{(\ell)}(\#r) \, \rangle = f^{(\ell)}, \qquad \qquad \le \ell = 1,2  \ri
 ,
\end{equation}
delivers the volume fraction of component material $\ell$, which is
subject to the constraint $\displaystyle{\sum^2_{\ell = 1}
f^{(\ell)}  = 1}$. The second statistical moment of $\Phi^{(\ell)}$
constitutes a two--point covariance function. Investigations
involving the electromagnetic SPFT have demonstrated that the
specific form of the covariance function has only a minor influence
on estimates of HCM constitutive parameters, for a range of
physically--plausible covariance functions \c{MLW01b}. Here we adopt
 the
physically--motivated form \cite{TKN82}
\begin{equation}
\langle \, \Phi^{(\ell)} (\#r) \, \Phi^{(\ell)} (\#r')\,\rangle =
\left\{
\begin{array}{lll}
\langle \, \Phi^{(\ell)} (\#r) \, \rangle \langle \Phi^{(\ell)}
(\#r')\,\rangle\,, & & \hspace{10mm}  | \, \=U^{-1}\cdot \le   \#r - \#r' \ri | > L \,\\ && \hspace{25mm} \\
\langle \, \Phi^{(\ell)} (\#r) \, \rangle \,, && \hspace{10mm}
 | \, \=U^{-1} \cdot \le  \#r -
\#r' \ri | \leq L\,
\end{array}
\right.,
 \l{cov}
\end{equation}
which has been widely used in   electromagnetic and elastodynamic
SPFT studies. The correlation length $L$ in eq. \r{cov} is required
to be much smaller than the associated piezoelectric wavelengths,
but larger than the particle size parameter $\eta$.

\subsection{Comparison material}

 A  homogeneous comparison material
 provides the initial ansatz for an
iterative procedure that delivers a succession of  SPFT estimates of
the HCM constitutive parameters \c{spft_zhuck1}. Accordingly, the
comparison material represents the lowest--order SPFT estimate of
the HCM. In consonance with  the component materials,  the
comparison material is an  orthorhombic $mm2$ piezoelectric
material, in general. The piezoelectric constitutive properties of
this orthorhombic comparison material (OCM) are  encapsulated by its
extended stiffness symbol $\breve{C}^{(ocm)}_{lMPq}$ (and its
12$\times$12 matrix equivalent $\breve{\*C}^{(ocm)}$)
  and
extended density symbol $\breve{\rho}_{MP}^{(ocm)}$ (and its
4$\times$4 matrix equivalent
$\mbox{\boldmath$\*{\breve{\rho}}$}^{(ocm)}$).

In order to establish the spectral Green function for the
OCM~---~which is a key element in the SPFT formulation~---~we first
consider the corresponding extended equation of motion. This
 may be written in the frequency domain as \c{Ma_Wang}
\begin{equation}
\breve{C}^{(ocm)}_{lMPq} \partial_l \partial_q \breve{u}_P +
\omega^2 \breve{u}_M = - \breve{F}_M,
\end{equation}
where the extended displacement
\begin{equation}
\breve{u}_M = \left\{ \begin{array}{lr} u_m, & M=m=1,2,3 \\
\Phi, & P=4
\end{array}
\right.
\end{equation}
combines the displacement $u_m$ and electric  scalar potential
$\Phi$, and the extended body force
\begin{equation}
\breve{F}_M = \left\{ \begin{array}{lr} F_m, & M=m=1,2,3 \\
-q, & M=4
\end{array}
\right.
\end{equation}
combines the body force $F_m$ and the electric charge $q$.
 Accordingly, the sought after  spectral Green function for the OCM emerges as
 the 4$\times$4 matrix
\begin{equation} \l{G_ocm_def}
\*G^{(ocm)}(\#k)=\big[k^2\*a(\#{\hat{k}})-\omega^2
\mbox{\boldmath$\*{\breve{\rho}}$}^{(ocm)} \;\big]^{-1},
\end{equation}
where the 4$\times$4 matrix $\*a(\#{\hat{k}})$  has entries
\begin{equation}
 \les \*a(\#{\hat{k}})\ris_{MP}= \displaystyle{\frac{k_s
\breve{C}_{sMPq}^{(ocm)} k_q}{k^2}} .
\end{equation}
Herein,  $\#k = k \#{\hat{k}}$  $ \equiv \le k_1, k_2, k_3 \ri$ with
$ \#{\hat{k}} = ( \sin \theta \cos \phi$, $\sin \theta \sin \phi$,
$\cos \theta )$.

The OCM extended constitutive symbols
 $\breve{C}^{(ocm)}_{lMPq}$ and $\breve{\rho}^{(ocm)}_{MP}$ are derived
via the imposition of the two conditions  \cite[eqs.
(2.72),(2.73)]{spft_zhuck1}
\begin{eqnarray}
&& \langle \Phi^{(1)} (\#r) \,\xi_{lMPq}^{(1)} + \Phi^{(2)} (\#r)\,
\xi_{lMPq}^{(2)} \rangle
 =0, \l{s1}\\
&&\langle \Phi^{(1)} (\#r) \les \breve{\rho}^{(1)}-
\breve{\rho}^{(ocm)} \ris_{MP} + \Phi^{(2)} (\#r) \les
\breve{\rho}^{(2)}- \breve{\rho}^{(ocm)} \ris_{MP} \rangle
 =0 ,\l{s2}
\end{eqnarray}
which is necessary  to remove certain secular terms. In eq.  \r{s1},
the quantities
\begin{equation}\label{xi_eq}
\xi_{lMPq}^{(\ell)}=\le \breve{C}_{lMSt}^{(\ell)} -
\breve{C}_{lMSt}^{(ocm)} \ri \eta_{StPq}^{(\ell)}, \qquad \quad
(\ell = 1,2),
\end{equation}
where $\eta^{(\ell)}_{StPq}$ is given implicitly through
\begin{eqnarray}
&&\breve{S}^{(\ell)}_{Pq} = \eta^{(\ell)}_{PqSt} f^{(\ell)}_{St}, \l{edef}\\
&&f^{(\ell)}_{Tj} = \breve{S}^{(\ell)}_{Tj}+W_{TjlM} \le
\breve{C}_{lMPq}^{(\ell)} - \breve{C}_{lMPq}^{(ocm)} \ri
\breve{S}^{(\ell)}_{Pq}, \l{fdef}
\end{eqnarray}
with the renormalization tensor
\begin{equation}\label{ellipse_Sint2}
W_{PstU} =\left\{
\begin{array}{lr}\displaystyle{
\frac{1}{8\pi} \int_{0}^{2\pi}d\phi \int_{0}^{\pi}d\theta \;
\frac{\sin \theta}{(\*U^{-1}\cdot \#{\hat{k}})\cdot(\*U^{-1}\cdot
\#{\hat{k}} )} \times }& \vspace{4pt}\\\displaystyle{(\*U^{-1}\cdot
\#{\hat{k}})_t \Big\{ (\*U^{-1}\cdot \#{\hat{k}})_s \big[\, \*a^{-1}
(\*U^{-1}\cdot \#{\hat{k}}) \, \big]_{pU}}
& \vspace{4pt}\\
\displaystyle{+(\*U^{-1}\cdot \#{\hat{k}})_p \big[ \, \*a^{-1}
(\*U^{-1}\cdot \#{\hat{k}})\,
\big]_{sU} \Big\}},&\textrm{P}=\textrm{p}=1,\,2,\,3\vspace{4pt}\\
&\\ \displaystyle{ \frac{1}{8\pi} \int_{0}^{2\pi}d\phi
\int_{0}^{\pi}d\theta \sin \theta \frac{(\*U^{-1}\cdot
\#{\hat{k}})_t (\*U^{-1}\cdot \#{\hat{k}})_s \big[\, \*a^{-1}
(\*U^{-1}\cdot \#{\hat{k}}) \, \big]_{pU}}{(\*U^{-1}\cdot
\#{\hat{k}})\cdot(\*U^{-1}\cdot \#{\hat{k}} )}},&\:\:\textrm{P}=4
\end{array}\right. .
\end{equation}

Upon substituting eqs. \r{xi_eq}--\r{fdef} into  eq. \r{s1},
exploiting eq. \r{vf}, and  after  some algebraic manipulations, we
obtain
\begin{equation}\label{fullxi}
f^{(1)}\les \le \breve{\*C}^{(1)}-\breve{\*C}^{(ocm)}
\ri^{\dagger}+\*W\, \ris^{\dagger} + f^{(2)}\les \le
\breve{\*C}^{(2)}-\breve{\*C}^{(ocm)} \ri^{\dagger}+\*W\,
\ris^{\dagger} = \*0_{\,12\times12},
\end{equation}
wherein the 12$\times$12 matrix equivalent
 of  $W_{RstU}$ (namely,  $\*W$) has been
 introduced and $^{\dagger}$ denotes the matrix  operation defined
in Appendix~A. The OCM stiffness matrix may be extracted from
\r{fullxi} as
\begin{equation}\label{1stnewit}
\breve{\*C}^{(ocm)}=\breve{\*C}^{(1)}+f^{(2)} \les\,
\mbox{\boldmath$\*\tau$}+(\breve{\*C}^{(2)}-\breve{\*C}^{(ocm)})\cdot
\*W \, \ris^{\dagger} \cdot\le
\breve{\*C}^{(1)}-\breve{\*C}^{(2)}\ri,
\end{equation}
where ${\mbox{\boldmath$\*\tau$}}$ is the $12\times 12$ matrix
representation of the extended identity $\tau_{rSTu} = \tau_{RstU}$,
as described in Appendix~A. By standard numerical procedures, such
as the Jacobi method \c{Bagnara}, the nonlinear relation
\r{1stnewit} is solved for $\breve{\*C}^{(ocm)}$.

After combining eq. \r{vf} with eq. \r{s2}, it follows immediately
that the OCM density is the volume average of the densities of the
component materials `1' and `2'; i.e.,
\begin{equation}
\mbox{\boldmath$\*{\breve{\rho}}$}^{(ocm)} =
f^{(1)}\mbox{\boldmath$\*{\breve{\rho}}$}^{(1)}+f^{(2)}\mbox{\boldmath$\*{\breve{\rho}}$}^{(2)}.
\end{equation}

\subsection{Second-order SPFT}\label{sec_SPFT}

Building upon the corresponding results for the elastodynamic SPFT
\c{spft_zhuck1},  the second--order\footnote{The first--order SPFT
estimate is identical to the zeroth--order SPFT estimate which is
represented by the comparison material.} estimates of the HCM
extended stiffness and density symbols may be expressed in terms of
 three--dimensional integrals as
\begin{eqnarray}\label{spft_int}
\breve{C}^{(spft)}_{lMPq} &=&
\breve{C}^{(ocm)}_{lMPq}-\frac{\omega^2}{2}\int d^3k \;
\frac{k_t}{k^2}\, B^{lMrs}_{tUPq}(\#k) \les \,
\mbox{\boldmath$\*{\breve{\rho}}$}^{(ocm)} \, \ris_{XY} \, \les \,
\*G^{(ocm)} (\#k)\,
\ris_{YU}\times\nonumber\\
&& \, \lec k_s \les \, \*a^{-1} (\#{\hat{k}})\, \ris_{rX}+k_r \les
\, \*a^{-1} (\#{\hat{k}}) \,\ris_{sX} \ric - \\
&&\frac{\omega^2}{2}\int d^3k \;  \frac{k_t}{k^2}\,
B^{lM4s}_{tUPq}(\#k) \les \,
\mbox{\boldmath$\*{\breve{\rho}}$}^{(ocm)} \, \ris_{XY} \, \les \,
\*G^{(ocm)} (\#k)\, \ris_{YU} \, \lec k_s \les \, \*a^{-1}
(\#{\hat{k}})\, \ris_{4X}\ric\nonumber
\end{eqnarray}
and
\begin{equation}\label{rho_int}
\breve{\rho}^{(spft)}_{MP}= \breve{\rho}^{(ocm)}_{MP}+\omega^2\int
d^3k \; B_{MSUP}(\#k) \les \*G^{(ocm)} (\#k) \ris_{SU}.
\end{equation}
The symbols $B^{lMRs}_{tUPq}(\#k)$ and $B_{MSUP}(\#k)$ represent the
spectral covariance functions given as
\begin{equation}\label{cov_fn}
\left.
\begin{array}{l}
\displaystyle{
 B^{lMNs}_{tUPq} (\#k) = \frac{ \le
\xi^{(2)}_{lMNs} - \xi^{(1)}_{lMNs} \ri \le \xi^{(2)}_{tUPq} -
\xi^{(1)}_{tUPq} \ri}{8 \pi^3} \int d^3 R \;
 \, \Gamma (\#R) \, \exp \le -i \#k \cdot  \#R
\ri} \vspace{4pt}\\
\displaystyle{ B_{MSUP}(\#k)= \frac{\le
\rho^{(2)}_{MS}-\rho^{(1)}_{MS}\ri\le
\rho^{(2)}_{UP}-\rho^{(1)}_{UP}\ri }{8 \pi^3} \int d^3R \; \Gamma
(\#R) \, \exp \le -i \#k \cdot  \#R \ri}
\end{array}
\right\},
\end{equation}
with
\begin{eqnarray}
\Gamma (\#R)= \Gamma (\#r - \#r') &=& \langle \, \Phi^{(1)}  (\#r)
\, \Phi^{(1)} (\#r')\,\rangle -  \langle \, \Phi^{(1)}  (\#r)
\,\rangle \, \langle
\, \Phi^{(1)} (\#r')\,\rangle\nonumber\\
&\equiv& \langle \, \Phi^{(2)} (\#r) \, \Phi^{(2)} (\#r')\,\rangle -
\langle \, \Phi^{(2)}  (\#r) \,\rangle \, \langle \, \Phi^{(2)}
(\#r')\,\rangle.
\end{eqnarray}

In order to make the integrals in the expressions for
$\breve{C}^{(spft)}_{lMPq}$ and $\breve{\rho}^{(spft)}_{MP}$
presented in eqs. \r{spft_int} and \r{rho_int}  numerically
tractable, we simplify them  as follows. Let us begin with the
integral on the right sides of eqs. \r{cov_fn}. Upon implementing
the step function--shaped covariance function \r{cov}, we find
\begin{equation}
\int d^3 R \;
 \, \Gamma (\#R) \, \exp \le -i \#k \cdot  \#R
\ri = \int_{|\#R | \leq L} d^3R \; \exp \les -i \le \*U \cdot \#k
\ri \cdot \#R \ris.
\end{equation}
Thereby, the expressions for $B^{lMRs}_{tUPq}(\#k)$ and
$B_{MSUP}(\#k)$ reduce to
\begin{equation}\label{covrfn}
\left.
\begin{array}{l}
\displaystyle{
 B^{lMRs}_{tUPq} (\#k) =  \frac{f^{(1)}f^{(2)}
 \le \xi^{(2)}_{lMRs} - \xi^{(1)}_{lMRs} \ri \le
\xi^{(2)}_{tUPq} - \xi^{(1)}_{tUPq} \ri }{2 \le \pi k \sigma \ri^2}
\les \frac{\sin \le k\sigma L \ri}{k\sigma} -L \cos \le k\sigma L
\ri
\ris} \vspace{4pt} \\
 \displaystyle{ B_{MSUP} (\#k) =  \frac{f^{(1)}f^{(2)}
\le \rho^{(2)}_{MS}-\rho^{(1)}_{MS}\ri \le
\rho^{(2)}_{UP}-\rho^{(1)}_{UP}\ri }{2 \le \pi k \sigma \ri^2} \les
\frac{\sin \le k\sigma L \ri}{k\sigma} -L \cos \le k\sigma L \ri
\ris}
\end{array}
\right\},
\end{equation}
wherein the scalar function
\begin{equation}
\sigma\equiv \sigma(\theta,\phi)=\sqrt{a^2 \sin^2\theta
\cos^2\phi+b^2 \sin^2\theta \sin^2\phi +c^2\cos^2\theta}
\end{equation}
is introduced.

Now we turn  to the integrals in \r{spft_int} and \r{rho_int}. In
analogy with the corresponding expression for the elastodynamic SPFT
\c{DML}, the spectral Green function $\*G^{(ocm)} (\#k)$ may be
conveniently expressed as
\begin{equation} \l{G_def}
\*G^{(ocm)} (\#k) =  \frac{\*D^{} (\#k)}{\Delta (\#k)}\,,
\end{equation}
where the 4$\times$4 matrix function
\begin{equation}
\*D (\#k) = \adj \, \les \, k^2 \, \*a(\#{\hat{k}}) -
\omega^2\mbox{\boldmath$\*{\breve{\rho}}$}^{(ocm)} \ris
\end{equation}
and the scalar function
\begin{eqnarray}\label{Delta_def}
\Delta (k) &=&k^8 \det\les \, \*a(\#{\hat{k}})\, \ris -\Tr\lec
\adj\les k^2\*a(\#{\hat{k}})\ris
\cdot\omega^2\mbox{\boldmath$\*{\breve{\rho}}$}^{(ocm)}\ric-
k^2 \, \Tr\les\adj(\omega^2\mbox{\boldmath$\*{\breve{\rho}}$}^{(ocm)})\cdot \*a(\#{\hat{k}})\ris\nonumber\\
& &+ k^4 \Big( \Tr\Big\{\les\*a(\#{\hat{k}})\ris_{44}\les \*a^\sharp
(\#{\hat{k}})\cdot
\adj(\omega^2\mbox{\boldmath$\*{{\breve{\rho}}}$}^\sharp )\ris
\Big\} -\les\*a(\#{\hat{k}})\ris_{41}\les\*a(\#{\hat{k}})\ris_{14}
 \les \adj(\omega^2\mbox{\boldmath$\*{{\breve{\rho}}
}$}^\sharp) \ris_{11} \nonumber\\
& & -\les\*a(\#{\hat{k}})\ris_{42}\les\*a(\#{\hat{k}})\ris_{24}
 \les \adj(\omega^2\mbox{\boldmath$\*{{\breve{\rho}} }$}^\sharp)  \ris_{22}
-\les\*a(\#{\hat{k}})\ris_{31}\les\*a(\#{\hat{k}})\ris_{13} \les
\adj(\omega^2\mbox{\boldmath$\*{{\breve{\rho}} }$}^\sharp) \ris_{33}
\Big),
\end{eqnarray}
with the 3$\times$3 matrixes $\*a^\sharp$ and
$\mbox{\boldmath$\*{{\breve{\rho}}}$}^\sharp$ having entries
\begin{equation}
\left. \begin{array}{l} \les \, \*a^\sharp \, \ris_{pq} =
  \les \, \*a(\#{\hat{k}}) \, \ris_{pq} \vspace{6pt}
  \\
\les \, \mbox{\boldmath$\*{{\breve{\rho}} }$}^\sharp \, \ris_{pq} =
  \les \, \mbox{\boldmath$\*{{\breve{\rho}} }$}^{(ocm)} \, \ris_{pq}
  \end{array}
  \right\}, \qquad (p,q = 1,2,3).
\end{equation}

Through exploiting eqs. \r{covrfn}  and \r{G_def}, the integrals in
eqs. \r{spft_int} and \r{rho_int} with respect to $k$ can be
evaluated by means of calculus of residues: The roots of
$\Delta(\#k) = 0$ give rise to seven poles in the complex--$k$
plane, located at  $k = 0, \pm p_1$, $ \pm p_2$, $ \pm p_3$, which
are chosen such that $ p_n$
 $ (n=1,2,3)$ lie in the upper half of the complex plane.
From eq. \r{Delta_def}, we find that the nonzero poles satisfy
\begin{eqnarray}
  p_1^2 &=& P_A-
\frac{1}{3} \le
  \frac{2^{1/3} P_B}{ P_C }- \frac{P_C}{  2^{1/3} }\ri,\\
  p_2^2 &=& P_A+ \frac{1}{3} \le \frac{(1+i\sqrt{3}) P_B}{ 2^{2/3} P_C }
  -\frac{(1-i\sqrt{3})P_C}{ 2^{4/3}}\ri, \\
  p_3^2 &=& P_A+
\frac{1}{3} \le
  \frac{(1-i\sqrt{3})P_B}{ 2^{2/3} P_C }
  -\frac{(1+i\sqrt{3})P_C}{2^{4/3} } \ri,
\end{eqnarray}
wherein
\begin{eqnarray}
  P_A &=& \frac{\omega^2 \Tr \lec \adj \les \,
\*a(\#{\hat{k}}) \, \ris \cdot \mbox{\boldmath$\*{\breve{\rho}}$}^{(ocm)} \ric }{3\,\det \les \, \*a(\#{\hat{k}}) \, \ris }, \\
  P_B &=& -C_A^2 + 3\,C_B,\\
  P_C &=& \les \, P_D + \le 4P_B^3+P_D^2 \ri^{1/2} \ris^{1/3},\\
  P_D &=& -2\,C_A^3 + 9\,C_A\,C_B-27\,C_C,
\end{eqnarray}
with
\begin{eqnarray}
C_A &=& \frac{-\omega^2 \, \Tr \lec \adj \les \,
\*a(\#{\hat{k}}) \, \ris \cdot \mbox{\boldmath$\*{\breve{\rho}}$}^{(ocm)} \ric }{\det \les \, \*a(\#{\hat{k}}) \, \ris }, \\
C_B &=& \frac{\omega^4}{\det \les \, \*a(\#{\hat{k}}) \, \ris
}\Bigg\{ \les \, \*a(\#{\hat{k}}) \, \ris_{44} \,  \Tr \,  \les \,
\*a^\sharp (\#{\hat{k}})\cdot \adj \, \le
\mbox{\boldmath$\*{{\breve{\rho}}}$}^\sharp\,\ri \,\ris +  \les \,
\*a(\#{\hat{k}}) \, \ris_{41}\les \, \*a(\#{\hat{k}}) \, \ris_{14}
\les \adj\,\le
\mbox{\boldmath$\*{\breve{\rho}}$}^{(ocm)}\,\ri \ris_{11} \nonumber\\
&& +  \les \, \*a(\#{\hat{k}}) \, \ris_{42}\les \, \*a(\#{\hat{k}})
\, \ris_{24} \les \adj\,\le
\mbox{\boldmath$\*{\breve{\rho}}$}^{(ocm)}\,\ri \ris_{22}  +  \les
\, \*a(\#{\hat{k}}) \, \ris_{43}\les \, \*a(\#{\hat{k}}) \,
\ris_{34} \les \adj\,\le
\mbox{\boldmath$\*{\breve{\rho}}$}^{(ocm)}\,\ri \ris_{33}
\Bigg\},\\
C_C &=& \frac{-\omega^6 \, \Tr \lec \adj \les \,
\mbox{\boldmath$\*{\breve{\rho}}$}^{(ocm)}\, \ris \cdot
\*a(\#{\hat{k}}) \ric } {\det \les \, \*a(\#{\hat{k}}) \, \ris }.
\end{eqnarray}
Thus, by application of the Cauchy residue theorem \c{Kwok},  the
SPFT estimates are delivered in terms of two--dimensional integrals
as
\begin{eqnarray}\label{final_full_int}
\breve{C}^{(spft)}_{lMPq} &= &
\breve{C}^{(ocm)}_{lMPq}+\frac{\omega^2 f^{(1)}f^{(2)}}{4\pi i
}\int_{\phi=0}^{2 \pi} \int_{\theta=0}^{\pi} d\phi \, d\theta \,
\frac{k_t  \sin\theta}{ \le k \sigma \ri^2 \; \det \les \,
\*a(\#{\hat{k}}) \, \ris } \les \,
\mbox{\boldmath$\*{\breve{\rho}}$}^{(ocm)} \, \ris_{XY} \les \,
\*b(\#{\hat{k}}) \, \ris_{YU} \nonumber\\
&& \times \bigg( \, \lec \xi^{(2)}_{lMrs} - \xi^{(1)}_{lMrs} \ric
\lec k_s \les \, \*a^{-1} (\#{\hat{k}})\, \ris_{rX}+k_r \les \,
\*a^{-1}
(\#{\hat{k}}) \,\ris_{sX} \ric \nonumber \\
&& + \lec \xi^{(2)}_{lm4s} - \xi^{(1)}_{lm4s} \ric \lec k_s \les \,
\*a^{-1} (\#{\hat{k}})\, \ris_{4X} \, \ric \bigg) \le
\xi^{(2)}_{tUPq} - \xi^{(1)}_{tUPq} \ri
\end{eqnarray}
and
\begin{equation} \l{rfinal_full_int}
\breve{\rho}^{(spft)}_{MP} =
\breve{\rho}^{(ocm)}_{MP}-\frac{\omega^2
 f^{(1)}f^{(2)}\le \breve{\rho}^{(2)}_{MS}-\breve{\rho}^{(1)}_{MS}\ri \le \breve{\rho}^{(2)}_{UP}-\breve{\rho}^{(1)}_{UP}\ri}{2\pi i
 } \int_{\phi=0}^{2 \pi}
 \int_{\theta=0}^{\pi} d\phi \,  d\theta
\,  \frac{\sin\theta}{\det \les \, \*a(\#{\hat{k}}) \, \ris}\, \les
\, \*b(\#{\hat{k}}) \, \ris_{SU} ,
\end{equation}
where the 4$\times$4 matrix
\begin{eqnarray}\label{Residue}
\*b(\#{\hat{k}})&=&\frac{1}{2i}\Bigg\{ \frac{e^{iL\sigma p_1}\*D(p_1
\*U \cdot \#{\hat{k}} )}{\sigma
p_1^4(p_1^2-p_2^2)(p_1^2-p_3^2)}\bigg(1-iL\sigma p_1\bigg) -
\frac{e^{iL\sigma p_2}\*D(p_2 \*U \cdot \#{\hat{k}} )}{\sigma
p_2^4(p_1^2-p_2^2)(p_2^2-p_3^2)}\bigg(1-iL\sigma p_2\bigg)\nonumber\\
& &+\frac{e^{iL\sigma p_3}\*D(p_3 \*U \cdot \#{\hat{k}} )}{\sigma
p_3^4(p_2^2-p_3^2)(p_1^2-p_3^2)}\bigg(1-iL\sigma p_3\bigg) \nonumber \\
& & - \frac{1}{\sigma
p_1^2p_2^2p_3^2}\bigg[\*D(\#0)\Big(\frac{1}{p_1^2}+\frac{1}{p_2^2}+\frac{1}{p_3^2}+\frac{\sigma^2L^2}{2}\Big)+
\frac{1}{2}  \frac{\partial^2}{\partial k^2 } \*D(\#0)\bigg]
\Bigg\}.
\end{eqnarray}
The expressions for the second--order SPFT estimates
$\breve{C}^{(spft)}_{lMPq}$ and $\breve{\rho}^{(spft)}_{MP}$ in eqs.
\r{final_full_int} and \r{rfinal_full_int} may be evaluated by
standard numerical methods \c{num_methods}.

It is particularly noteworthy that  $\breve{C}^{(spft)}_{lMPq}$ and
$\breve{\rho}^{(spft)}_{MP}$ are complex--valued for $L > 0$, even
when the corresponding quantities for the
 component materials, i.e.,
$\breve{C}^{(\ell)}_{lMPq}$ and $\breve{\rho}^{(\ell)}_{MP}$  ($\ell
= 1,2$), are real--valued.  This reflects the fact that the SPFT
accommodates losses due to scattering \c{spft_form}. From energy
considerations, the imaginary part of the extended  compliance
matrix, namely \c{Auld_book}
\begin{equation} \l{M_def}
\breve{\*M}^{(spft))} = \le \begin{array}{ccc} \le \*C^{(spft)}
\ri^{-1} & & \le \*C^{(spft)} \ri^{-1} \cdot \*e^{(spft)} \\
 \les  \le \*C^{(spft)} \ri^{-1} \cdot \*e^{(spft)} \ris^{T} && \mbox{\boldmath$\*\epsilon$}^{(spft)} + \le \*e^{(spft)} \ri^T
\cdot  \le \*C^{(spft)} \ri^{-1}  \cdot \*e^{(spft)}
\end{array}
\ri,
\end{equation}
is required to
 be
positive definite for passive materials \cite{Holland_Comploss}. The
constitutive matrixes $\*C^{(spft)}$, $\*e^{(spft)}$ and
$\mbox{\boldmath$\*\epsilon$}^{(spft)}$ on the right side of eq.
\r{M_def} are related to the extended stiffness matrix
$\*{\breve{C}}^{(spft)}$ (and thereby to the extended stiffness
symbol $\breve{C}^{(spft)}_{lMPq}$)  per eq. \r{stiffness_mat_def}.

\section{Numerical results}

\subsection{Preliminaries} \l{NR_Prelim}

In order to  illustrate the theory presented in
\S\ref{Theory_section}, let us now consider a  representative
numerical example. A comparison for the SPFT estimate of the HCM
constitutive parameters is provided by the corresponding results
computed using the Mori--Tanaka  formalism
\cite{Mori-Tanaka,DunnTaya2,Mura,Lakhjcm}. In the case of
orthorhombic $mm2$ piezoelectric component materials, the
Mori-Tanaka estimate of the extended stiffness matrix for the HCM is
given by \cite{Odegard}
\begin{equation}
\breve{\*C}^{(MT)} = \breve{\*C}^{(1)} + f^{(2)} \le
\breve{\*C}^{(2)}-\breve{\*C}^{(1)} \ri\cdot \*B^{(MT)} \cdot \les
f^{(1)}\mbox{\boldmath$\*\tau$} + f^{(2)}\*B^{(MT)} \ris^{\dagger},
\end{equation}
where the 12$\times$12 matrix
\begin{equation}
\*B^{(MT)} = \les \mbox{\boldmath$\*\tau$} +
\*S^{(Esh)}\cdot\le\breve{\*C}^{(1)}\ri^{\dagger}\cdot \le
\breve{\*C}^{(2)}-\breve{\*C}^{(1)} \ri\ris ^{\dagger},
\end{equation}
with $\*S^{(Esh)}$ being the $12\times 12$ matrix representation of
the Eshelby tensor \cite{Eshelby,DunnTaya,Numerical_Eshelby}.
Details on evaluating $\*S^{(Esh)}$ can be found in Appendix~B.

In the following, we present the numerical evaluation of the
12$\times$12 extended stiffness matrix of the HCM, namely
$\breve{\*C}^{(hcm)}$, as estimated by the lowest--order SPFT (i.e.,
$hcm =ocm$), the second--order SPFT (i.e., $hcm =spft$) and the
Mori--Tanaka  formalism (i.e., $hcm =MT$). The matrix
$\breve{\*C}^{(hcm)}$ has the form represented in eq.
\r{stiffness_mat_def}.  The second--order SPFT density tensor
$\rho^{(spft)}_{MP}$ is also evaluated; the numerical evaluation of
the lowest--order SPFT density $\rho_{MP}^{(ocm)}$ need not be
presented here as this quantity is simply the volume average of the
densities of the component materials.
 An angular frequency of $\omega=2\pi\times 10^6$ $\mbox{s}^{-1}$
 was selected
 for all second--order SPFT computations.

The eccentricities of the ellipsoidal component particles are
specified by the shape parameters $\lec a, b, c \ric$,  per eqs.
\r{r_shape} and \r{U_shape}. To allow direct comparison with results
from previous studies \cite{Odegard}, component material `1' was
taken to be the piezoelectric material polyvinylidene fluoride
(PVDF) while component material `2' was taken to be the
thermoplastic polyimide LaRC-SI, which has no piezoelectric
properties.
  The stiffness constitutive parameters of the component
materials are tabulated in Table~\ref{comp_prop_table1}. The nonzero
piezoelectric constitutive parameters of PVDF are: $e_{113} \equiv
e_{31} = 0.024$, $e_{223} \equiv e_{32} = 0.001$ and $e_{333} \equiv
e_{33} = -0.027$ in units of C m$^{-2}$. The dielectric constitutive
parameters of PVDF are: $\eps_{11} = 7.4$, $\eps_{22} = 9.6$ and
$\eps_{33} =7.6$, whereas those of LaRC-SI are: $\eps_{11} =
\eps_{22} = \eps_{33} = 2.8$, all in units of $\epso = 8.854 \times
10^{-12}$ F m${}^{-1}$ (the permittivity of free space). Lastly, the
densities of PVDF and LaRC-SI are 1750  and 1376, respectively, in
units of kg m$^{-3}$.

\begin{table}[h!]
  \centering
\begin{tabular}{|c|c|c|}
  \hline
  Stiffness parameter  & PVDF (GPa)  & LaRC-SI  (GPa)\\
  \hline
  $C^{}_{1111} \equiv C_{11}$    & 3.8   & 8.1 \\
  $C^{}_{1122} \equiv C_{12} $    & 1.9   & 5.4 \\
  $C^{}_{1133} \equiv C_{13}$    & 1.0   & 5.4 \\
  $C^{}_{2222} \equiv C_{22} $    & 3.2   & 8.1 \\
  $C^{}_{2233} \equiv C_{23}$    & 0.9   & 5.4 \\
  $C^{}_{3333} \equiv C_{33}$    & 1.2   & 8.1 \\
  $C^{}_{2323} \equiv C_{44} $    & 0.7   & 1.4   \\
  $C^{}_{1313} \equiv C_{55}$    & 0.9   & 1.4 \\
  $C^{}_{1212} \equiv C_{66} $    & 0.9   & 1.4 \\
  \hline
\end{tabular}
 \caption{The stiffness constitutive parameters of  the component materials in units of GPa (after \cite{Odegard}).}\label{comp_prop_table1}
\end{table}

\subsection{Lowest--order SPFT} \l{lo_SPFT}

\begin{figure}[h!]
\begin{center}
\resizebox{3.5in}{!}{\includegraphics{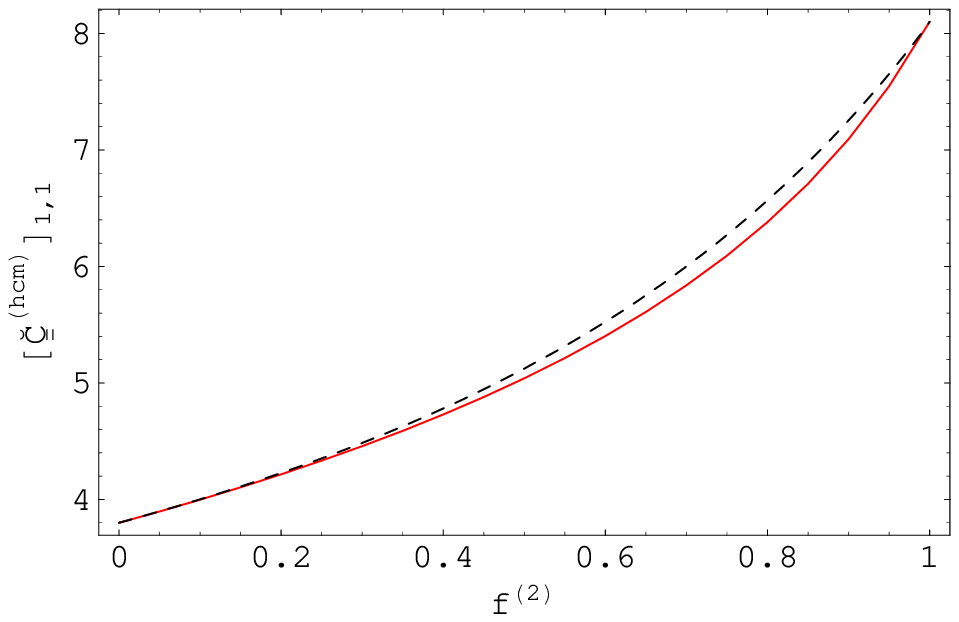}}
\resizebox{3.5in}{!}{\includegraphics{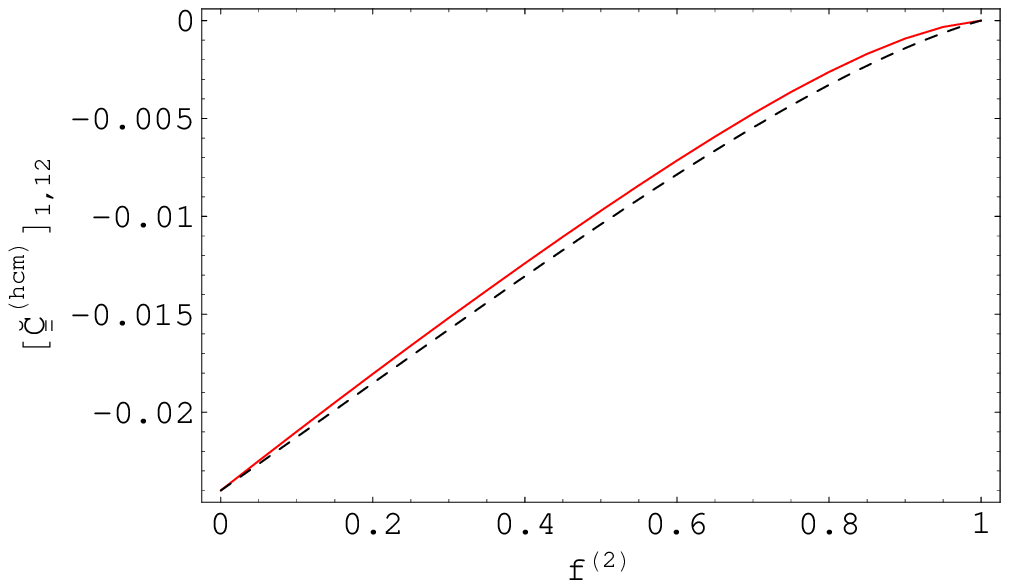}}
\resizebox{3.5in}{!}{\includegraphics{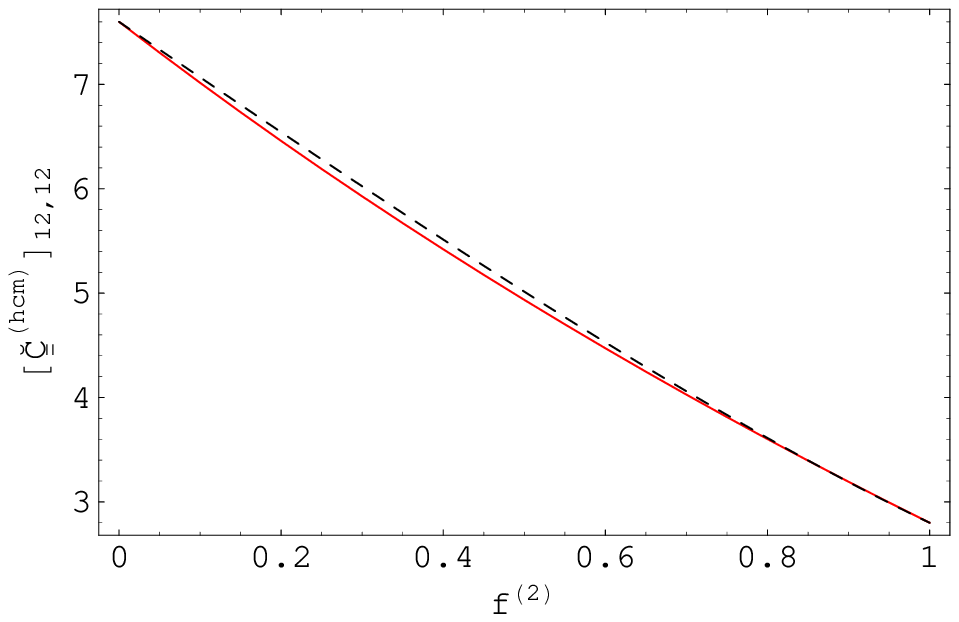}}
\caption{Plots of $\les \, \breve{\*C}^{(hcm)} \, \ris_{1,1}$ (in
GPa), $\les \, \breve{\*C}^{(hcm)} \, \ris_{1,12}$ (in C m$^{-2}$)
and $ (1/\epso) \les \, \breve{\*C}^{(hcm)} \, \ris_{12,12}$
 as estimated using the lowest--order SPFT (i.e.,
$hcm = ocm$) (black, dashed curves) and the Mori--Tanaka  formalism
(i.e., $hcm = MT$) (red, solid curves), versus the volume fraction
of component material `2'. Component material `1' is PVDF and
component material `2' is LaRC-SI, as described in
\S\ref{NR_Prelim}. The component materials are distributed as
spheres (i.e., $a=b=c$). }\label{mtspher1}
\end{center}
\end{figure}

We begin by considering the lowest--order SPFT estimates of the HCM
constitutive parameters. In  Fig.~\ref{mtspher1},  components of the
HCM extended stiffness matrix $\breve{\*C}^{(hcm)}$, as computed
using the lowest--order SPFT and the Mori--Tanaka formalism, are
plotted as functions of volume fraction $f^{(2)}$ for the case where
the component particles are spherical (i.e.,  $a=b=c$). Plots of
only a representative selection of the components of
$\breve{\*C}^{(hcm)}$ are presented in Fig.~\ref{mtspher1}; plots
for those components which are not presented in Fig.~\ref{mtspher1}
are qualitatively similar to those that are presented. Only
relatively minor differences between the lowest--order SPFT
estimates and the Mori--Tanaka estimates are observed, with the
differences between the two being greatest for mid--range values of
$f^{(2)}$. Plots for both the SPFT and Mori--Tanaka estimates are
necessarily constrained by the limits
\begin{equation}
\lim_{f^{(2)} \to 0}  \breve{\*C}^{(hcm)}
 =  \breve{\*C}^{(1)},\qquad
\lim_{f^{(2)} \to 1} \breve{\*C}^{(hcm)}  = \breve{\*C}^{(2)}.
\end{equation}

\begin{figure}[h!]
\begin{center}
\resizebox{3.5in}{!}{\includegraphics{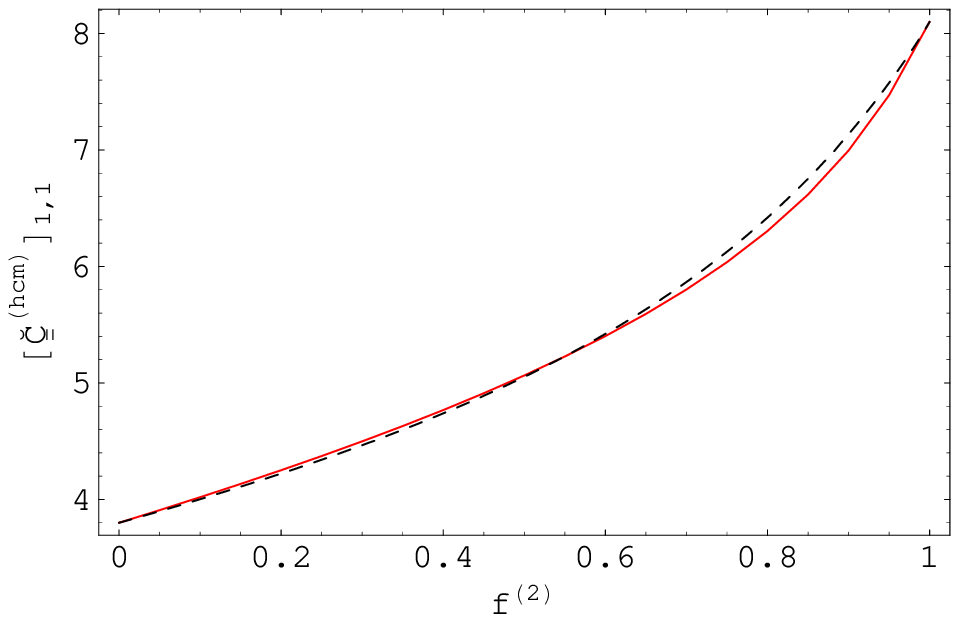}}
\resizebox{3.5in}{!}{\includegraphics{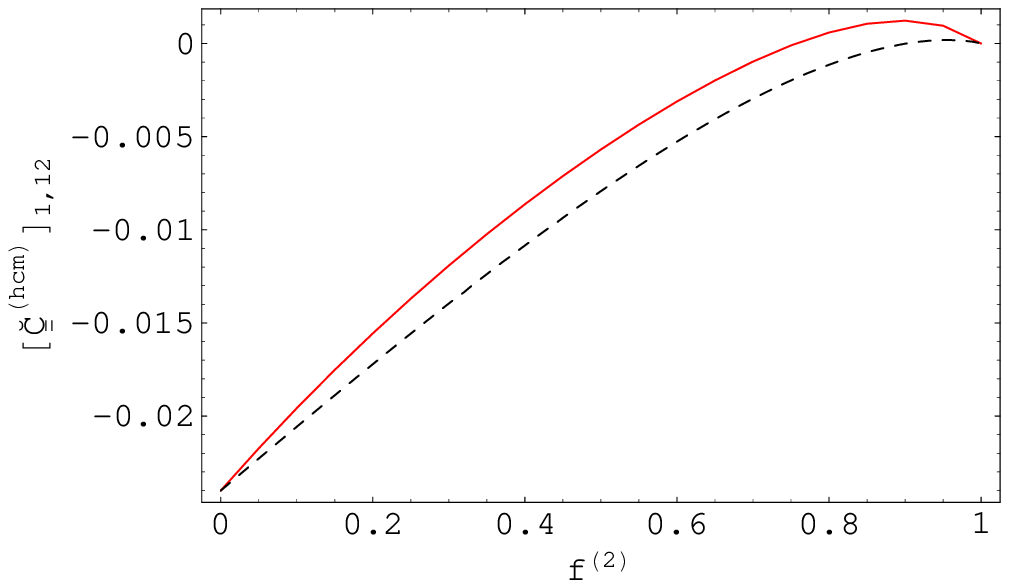}}
\resizebox{3.5in}{!}{\includegraphics{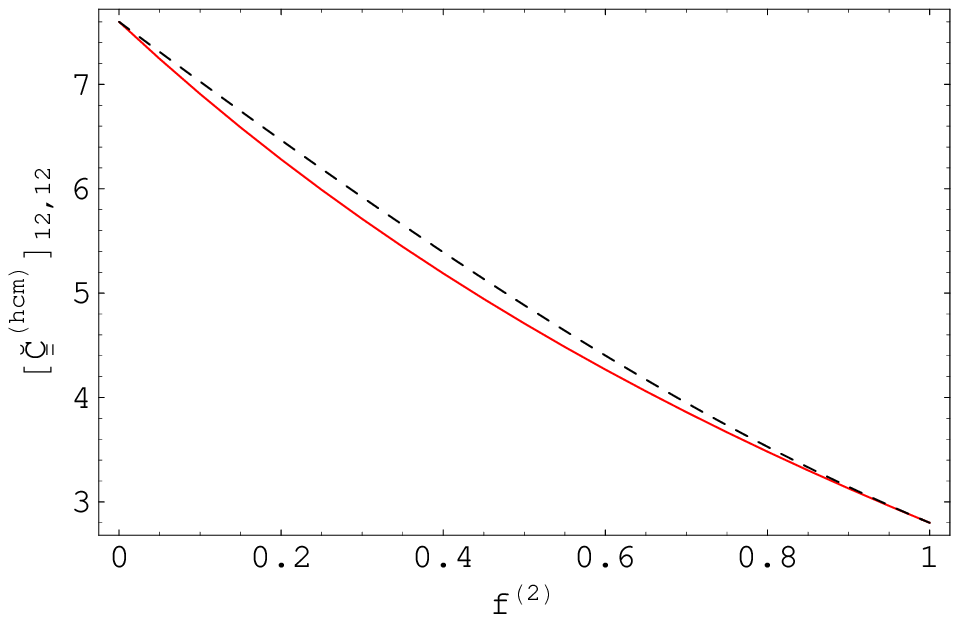}}
\caption{As Fig.~\ref{mtspher1} but with the component materials
distributed as ellipsoids with ($a/c=5$ and
$b/c=1.5$).}\label{eps1_1}
\end{center}
\end{figure}

\begin{figure}[h!]
\begin{center}
\resizebox{3.5in}{!}{\includegraphics{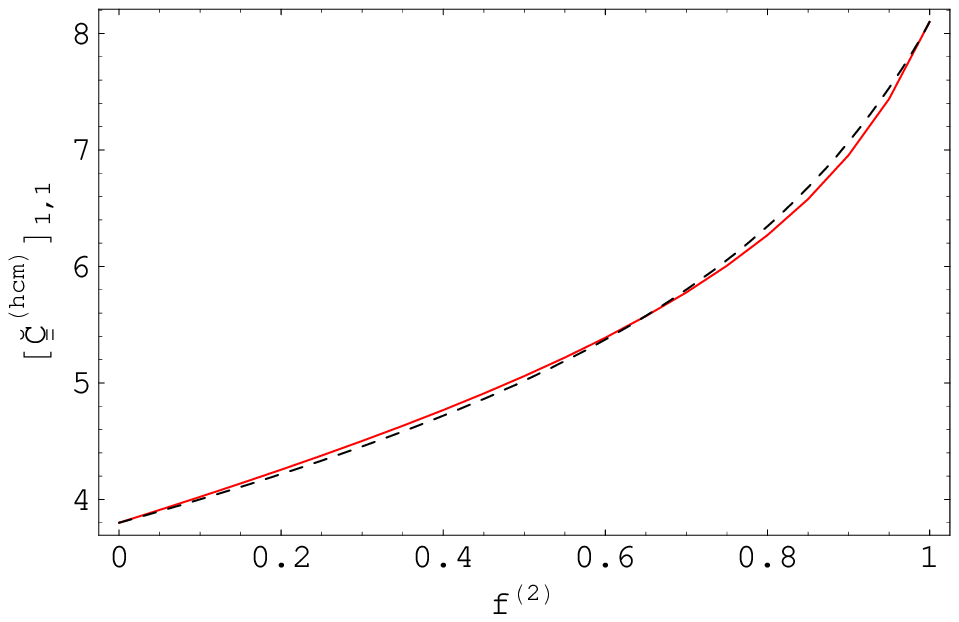}}
\resizebox{3.5in}{!}{\includegraphics{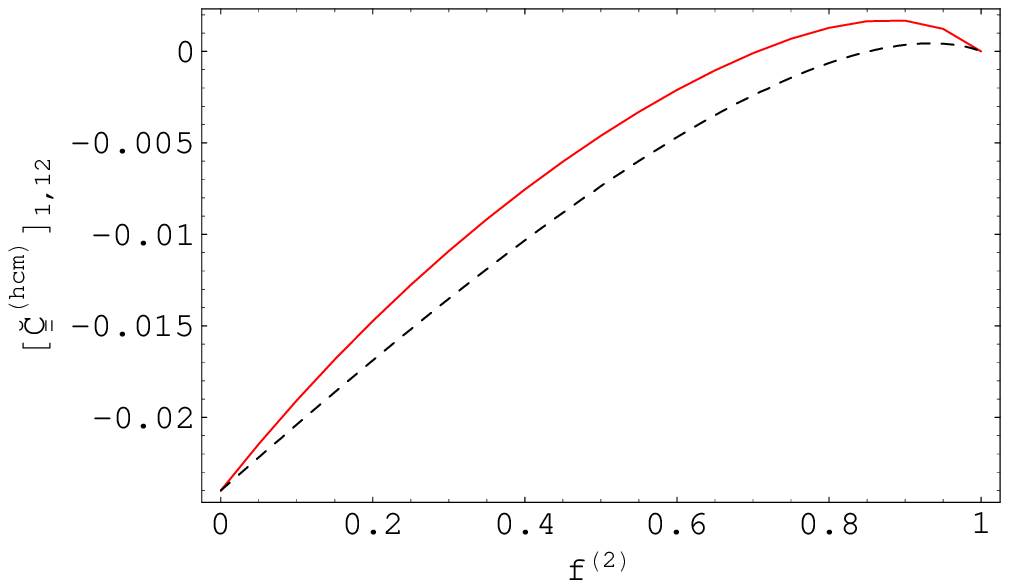}}
\resizebox{3.5in}{!}{\includegraphics{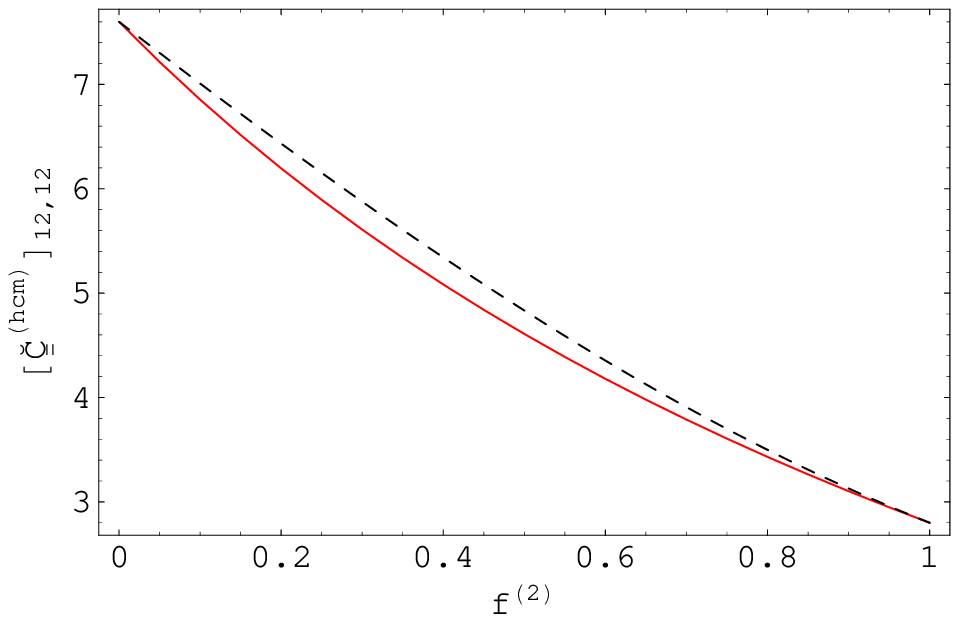}}
\caption{As Fig.~\ref{mtspher1} but with the component materials
distributed as ellipsoids with ($a/c=10$ and
$b/c=2$).}\label{eps2_1}
\end{center}
\end{figure}

 The corresponding graphs
for the  cases where the components particles are described by the
shape parameters
 $\lec a/c = 5, \, b/c = 1.5 \ric$ and $\lec a/c = 10, \, b/c = 2 \ric$
 are
provided in  Figs.~\ref{eps1_1} and \ref{eps2_1}, respectively. A
comparison of Figs.~\ref{mtspher1}--\ref{eps2_1} reveals that the
differences between the lowest--order SPFT and Mori--Tanaka
estimates are accentuated as the component particles become more
eccentric in shape, especially at mid--range values of $f^{(2)}$ for
the piezoelectric parameters and the dielectric parameters.

\subsection{Second--order SPFT estimate}
\begin{figure}[h!]
\begin{center}
\resizebox{2.9in}{!}{\includegraphics{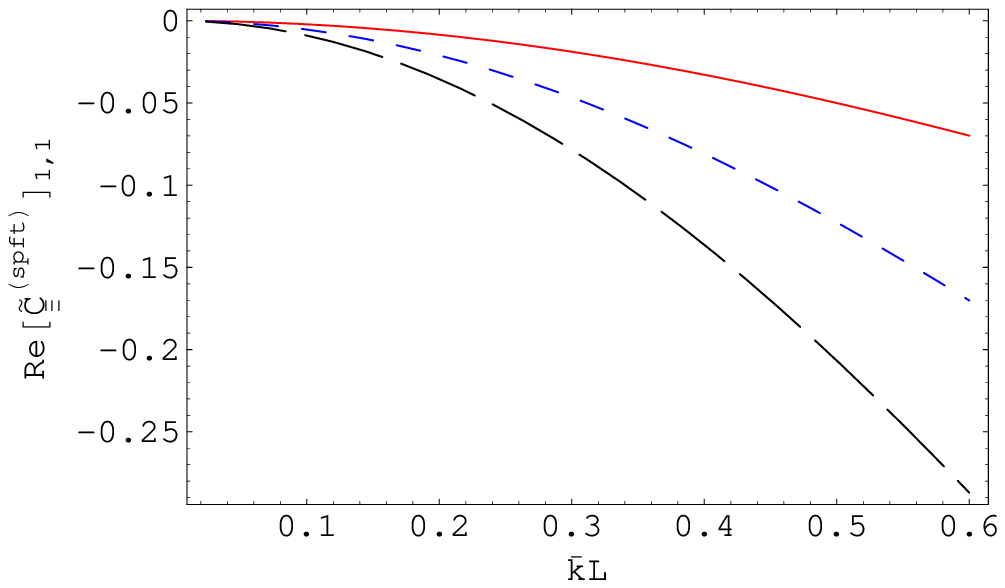}}
\resizebox{2.9in}{!}{\includegraphics{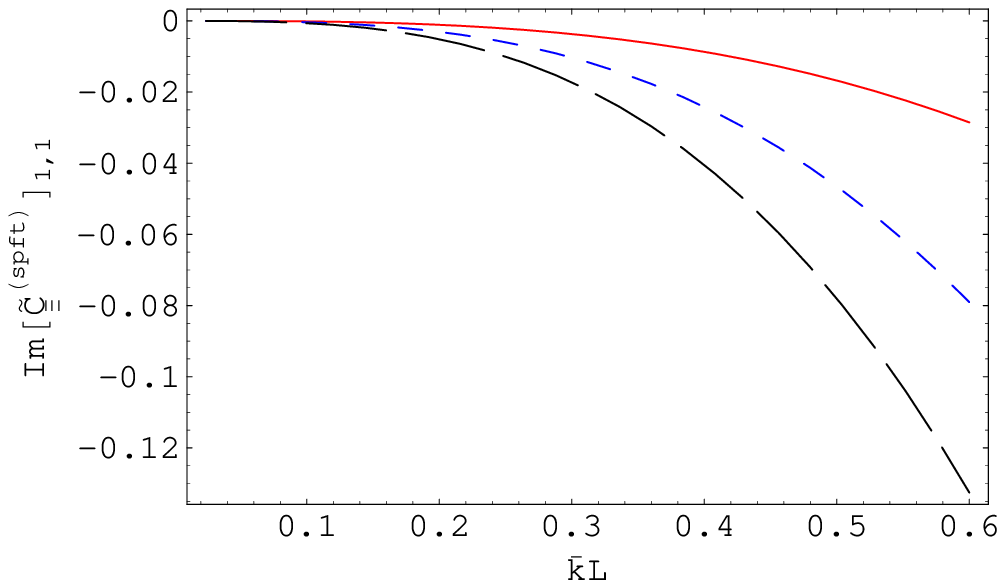}}\\
\resizebox{2.9in}{!}{\includegraphics{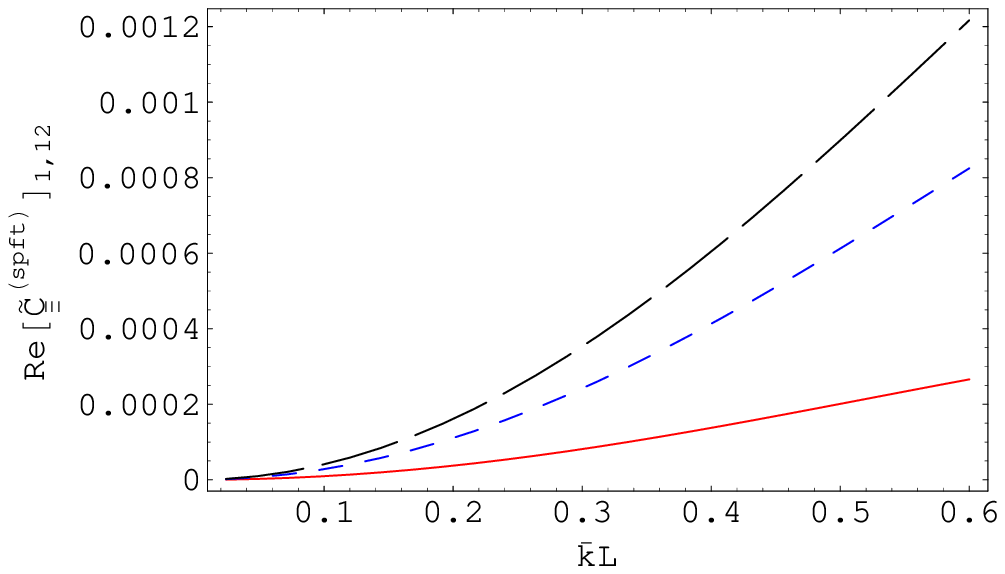}}
\resizebox{2.9in}{!}{\includegraphics{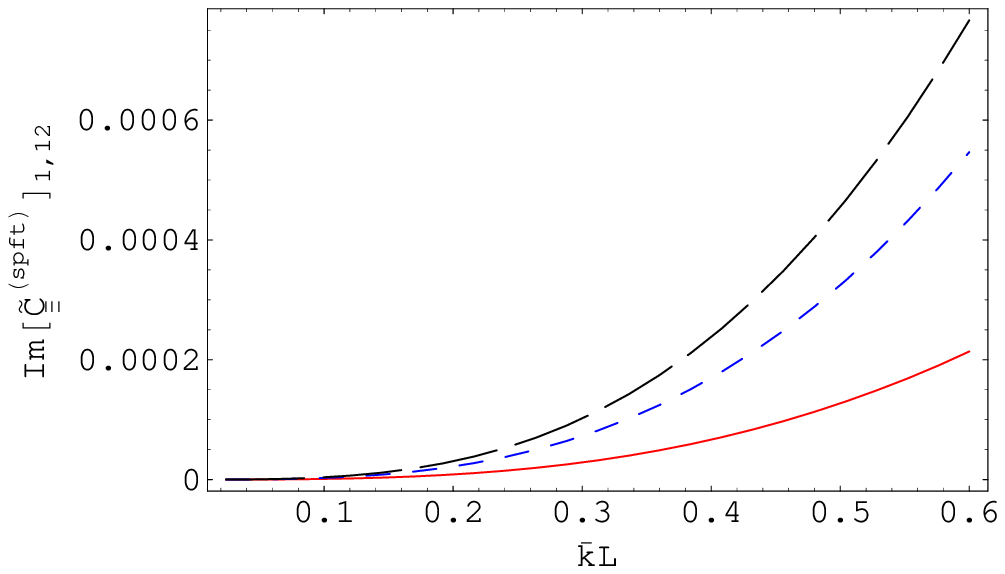}}\\
\resizebox{2.9in}{!}{\includegraphics{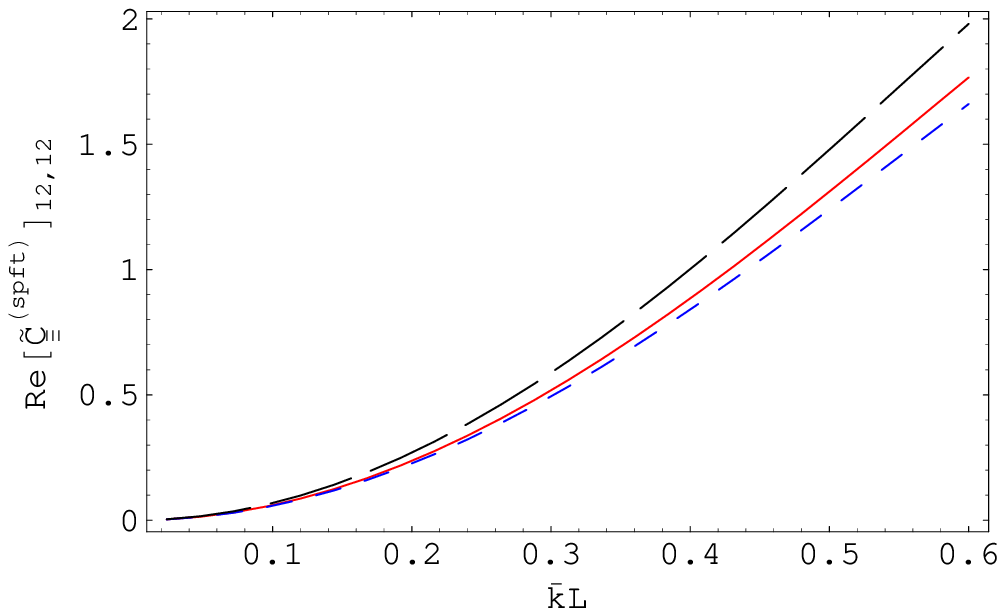}}
\resizebox{2.9in}{!}{\includegraphics{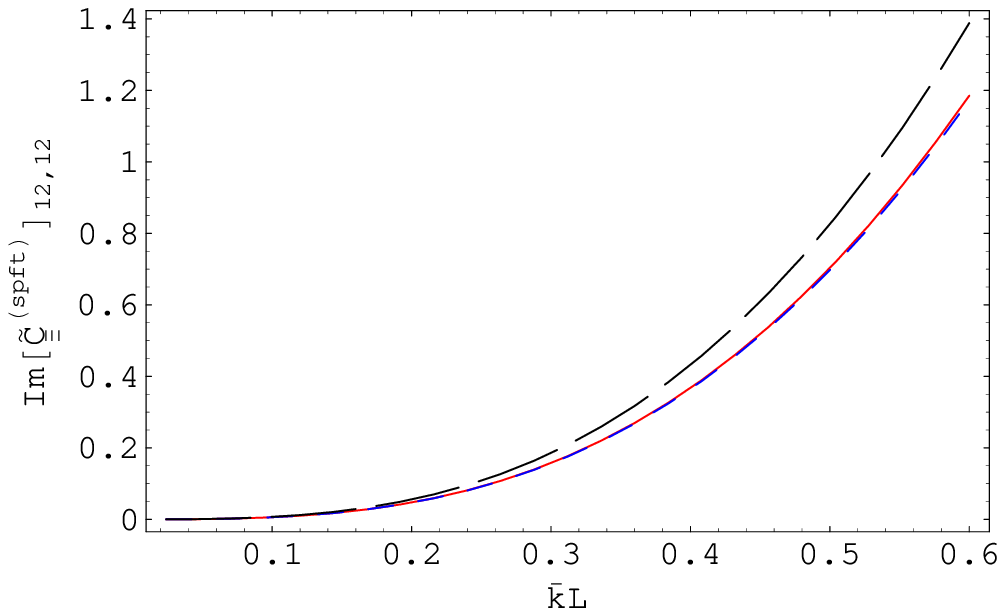}}
\caption{Plots of the real and imaginary parts of the second--order
SPFT estimates  $\les \, \tilde{\*C}^{(spft)} \, \ris_{1,1}$ (in
GPa), $\les \, \tilde{\*C}^{(spft)} \, \ris_{1,12}$ (in C m$^{-2}$)
and $ (10^3/\epso) \les \, \tilde{\*C}^{(spft)} \, \ris_{12,12}$,
where $ \tilde{\*C}^{(spft)} = \breve{\*C}^{(spft)} -
\breve{\*C}^{(ocm)} $, versus $\bar{k} L$, with $f^{(2)}=0.5$. The
results from the spherical particle (i.e., $a=b=c=1$) case (red,
solid line) are plotted alongside the cases with elliptical
particles with  $a=5$, $b=1.5$, $c=1$ (blue, short-dashed line) and
$a=10$, $b=2$, $c=1$ (black, long-dashed line). Component material
`1' is PVDF and component material `2' is LaRC-SI, as described in
\S\ref{NR_Prelim}.}\label{replotelas}
\end{center}
\end{figure}

Now let us turn to the second--order SPFT estimates of the HCM
constitutive parameters. We considered these quantities as functions
of $\bar{k} L$, where $\bar{k}$ is an approximate upper bound on the
 wavenumbers supported by the HCM, as estimated by \c{DML}
\begin{equation}
\bar{k} = \frac{\omega}{2} \le
\sqrt{\frac{\bar{\rho}^{}}{\bar{\lambda}^{} + 2 \bar{\mu}^{}}} +
\sqrt{\frac{\bar{\rho}^{}}{\bar{\mu}^{}} }\ri,
\end{equation}
wherein
\begin{equation}
\left.
\begin{array}{l}
\bar{\lambda}^{} = \displaystyle{\frac{1}{6}  \sum^2_{\ell = 1} \le
\, \left| \les\,\*C^{(\ell)} \, \ris_{12} \right| +   \left|
\les\,\*C^{(\ell)} \, \ris_{13} \right| +  \left| \les\,\*C^{(\ell)}
\,
\ris_{23} \right| \, \ri } \vspace{6pt} \\
\bar{\mu}^{} = \displaystyle{\frac{1}{6}  \sum^2_{\ell = 1} \le \,
\left| \les\,\*C^{(\ell)} \, \ris_{44} \right| +   \left|
\les\,\*C^{(\ell)} \, \ris_{55} \right| +  \left| \les\,\*C^{(\ell)}
\, \ris_{66} \right| \, \ri } \vspace{6pt} \\
\bar{\rho} = \displaystyle{\frac{1}{2} \,  \sum^2_{\ell = 1}
\rho^{(\ell)} }
\end{array}
\right\};
\end{equation}
and $L$ is the correlation length associated with the the two--point
covariance function \r{cov}. In Fig.~\ref{replotelas}, the real and
imaginary parts  of the components of
$\*{\tilde{C}}^{(spft)}=\breve{\*C}^{(spft)}-\breve{\*C}^{(ocm)}$
are plotted against $\bar{k} L$ for $f^{(2)} = 0.5$. The values of
the shape parameters $\lec a, \, b, \, c \ric$ correspond to those
used in the calculations for Figs.~\ref{mtspher1}--\ref{eps2_1}. As
in \S\ref{lo_SPFT}, only a representative selection of the
components of $\tilde{\*C}^{(spft)}$ are plotted in
Fig.~\ref{replotelas}; the graphs for those components that are not
represented in Fig.~\ref{replotelas} are qualitatively similar to
the graphs which do appear.

The second--order  corrections to the lowest--order SPFT estimates
are observed in Fig.~\ref{replotelas} to grow exponentially in
magnitude as the correlation length increases from zero.
Furthermore, the magnitudes of both the real and imaginary parts of
$\breve{\*C}^{(spft)}$ generally grow faster with increasingly
correlation length when the components particles are more eccentric
in shape.
 At $L=0$,
the second--order and lowest--order SPFT estimates coincide. While
the second--order corrections are relatively small compared to the
lowest--order SPFT estimates, a highly significant feature of the
second--order corrections is that these are complex--valued with
nonzero imaginary parts, even though  $ \breve{\*C}^{(a,b)} $ and $
\breve{\*C}^{(ocm)} $ are purely real--valued. We note that for all
 computations the imaginary part of the extended compliance matrix
$ \breve{\*M}^{(spft)}$ was found to be positive definite, which
corresponds to positive loss \c{Holland_Comploss}. Thus, the
emergence of nonzero imaginary parts of $\breve{\*C}^{(spft)} $
indicates that the HCM has acquired a dissipative nature, despite
the component materials being nondissipative. The dissipation is
attributed to scattering losses, since the second--order SPFT takes
into account interactions between spatially--distinct scattering
particles via the two--point covariance function \r{cov}. As the
correlation length increases, the number of scattering particles
that can mutually interact also increases, thereby increasing the
scattering loss per unit volume.

\begin{figure}[h!]
\begin{center}
\resizebox{3.5in}{!}{\includegraphics{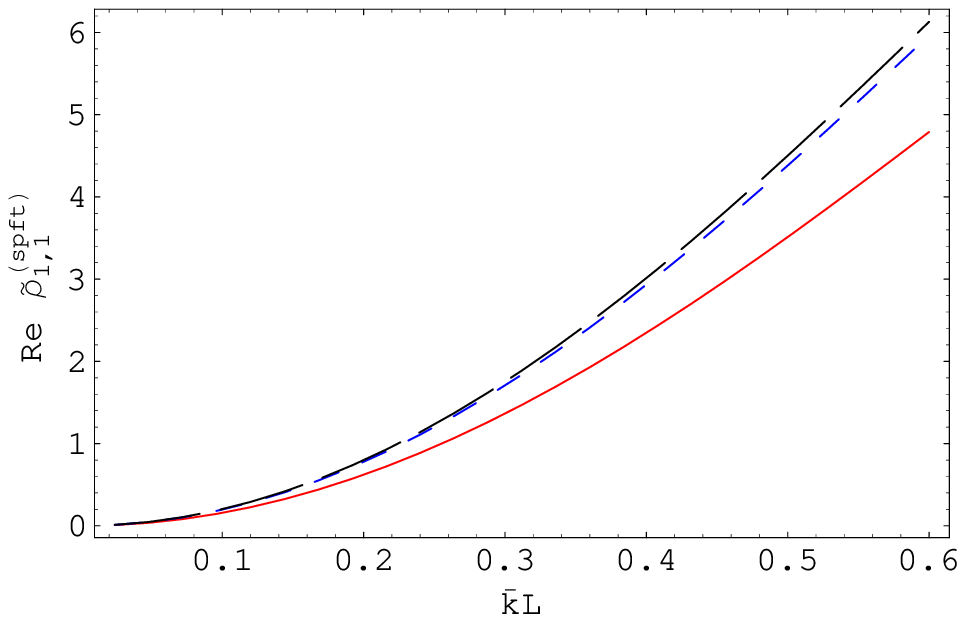}}
\resizebox{3.5in}{!}{\includegraphics{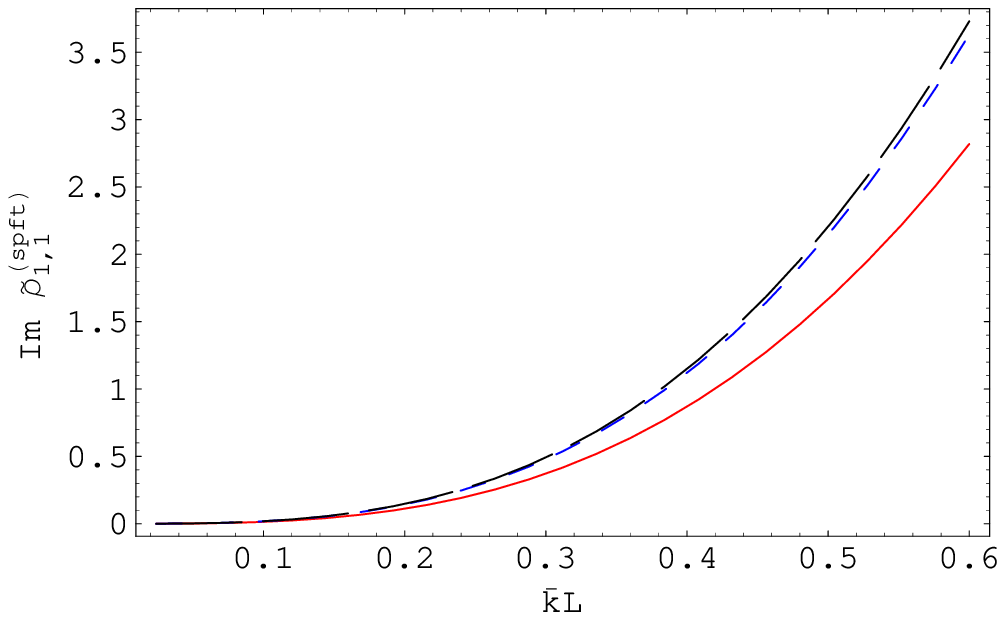}} \caption{ As
Fig.~\ref{replotelas} but with the real and imaginary parts of $\les
\, \mbox{\boldmath$\*{\tilde{\rho}}$}^{(spft)} \, \ris_{11}$ (in kg
m${}^{-3}$), where $\mbox{\boldmath$\*{\tilde{\rho}}$}^{(spft)} =
\mbox{\boldmath$\*{\breve{\rho}}$}^{(spft)} -
\mbox{\boldmath$\*{\breve{\rho}}$}^{(ocm)}$, plotted as functions of
$\bar{k} L$, with $f^{(2)}=0.5$. }\label{replotrho}
\end{center}
\end{figure}

Finally, we turn to the second--order SPFT estimate of the HCM
density. The  real and imaginary  parts  of the matrix entry $\les
\, \mbox{\boldmath$\*{\tilde{\rho}}$}^{(spft)}\, \ris_{11}$, wherein
$\mbox{\boldmath$\*{\tilde{\rho}}$}^{(spft)} =
\mbox{\boldmath$\*{\breve{\rho}}$}^{(spft)} -
\mbox{\boldmath$\*{\breve{\rho}}$}^{(ocm)}$,
 are plotted as functions of
$\bar{k} L$ in Fig.~\ref{replotrho}. The corresponding graphs for
$\les \, \mbox{\boldmath$\*{\tilde{\rho}}$}^{(spft)}\, \ris_{22}$
and $\les \, \mbox{\boldmath$\*{\tilde{\rho}}$}^{(spft)}\,
\ris_{33}$ are  much the same as those for $\les \,
\mbox{\boldmath$\*{\tilde{\rho}}$}^{(spft)}\, \ris_{11}$ but with
minor differences in magnitudes. The second--order SPFT estimates of
the HCM density exhibit
  characteristics similar to those of the corresponding HCM
  stiffness, piezoelectric and dielectric constitutive parameters.
  That is,
$
 \displaystyle{\lim_{L
\to 0} \rho^{(spft)}_{aa} =  \rho^{(ocm)}} $ and $ \Big\vert
\tilde{\rho}^{(spft)}_{aa} \Big\vert
 \ll \Big\vert \rho^{(ocm)} \Big\vert$
for $a = 1,2$ and $ 3$. Also, the differences between
$\mbox{\boldmath$\*{\breve{\rho}}$}^{(spft)}$ and
$\mbox{\boldmath$\*{\breve{\rho}}$}^{(ocm)}$ increase exponentially
as the correlation length increases, and this effect is most
accentuated when the component particles are most eccentric in
shape.

We remark that a complex--valued,  anisotropic density also crops up
in the second--order elastodynamic SPFT for orthotropic HCMs
\c{DML}, as well as in other homogenization scenarios
\c{Willis85,Milton_NJP2007}.

\section{Closing remarks}

The  linear SPFT has been fully developed for the case of
orthorhombic $mm2$ piezoelectric HCMs, based on component materials
distributed as oriented ellipsoidal particles.  The
multifunctionality of such HCMs  is central to the notion of
 metamaterials \c{Walser}.
The second--order estimates of the HCM constitutive parameters are
expressed in terms of numerically--tractable two--dimensional
integrals, for a specific choice of two--point covariance function.
This theoretical result further extends the application of the SPFT
in the homogenization of complex composites, effectively bridging
the elastodynamic SPFT for orthotropic HCMs \c{spft_zhuck1,DML} and
the electromagnetic SPFT for anisotropic dielectric HCMs
\c{Genchev,Z94}. Furthermore, the path has now been cleared towards
the development of the SPFT for piezoelectric/piezomagnetic HCMs
\c{Nan}, with bianisotropic electromagnetic properties
\c{spft_form}. Let us  remark that   the mathematical description of
piezoelectric HCMs presented herein  also extends to electrokinetic
processes \c{Adler}.

 From our theoretical considerations and representative
numerical studies,  the following conclusions were drawn:
\begin{itemize}

\item The lowest--order SPFT estimate of the stiffness,
piezoelectric and dielectric properties of the HCM are qualitatively
similar to those estimates provided by the Mori--Tanaka  formalism.

\item Differences between the estimates of
the lowest--order SPFT and the Mori--Tanaka  formalism are greatest
at mid--range values of the volume fraction, and
 accentuated when the  component particles are
eccentric in shape.

\item The second--order SPFT provides a  correction to the
lowest--order estimate of the HCM constitutive properties. The
magnitude of this correction is
 generally
larger when the  component particles are more eccentric in shape,
and
 vanishes as the correlation length tends to
zero.

\item While the correction provided
by the second--order SPFT is  relatively small in magnitude,  it is
highly significant as it indicates dissipation due to scattering
loss.

\end{itemize}

\section*{Appendix A}

The extended symbol $\breve{A}_{aMPq}$  ($a,q \in \lec 1,2,3 \ric$,
$M,P \in \lec 1,2,3,4 \ric$) may be conveniently  represented by the
 $12\times 12$ matrix with entries $\les \breve{\*A}
\ris_{\gamma \kappa}$  ($\gamma, \kappa \in \les 1, 12 \ris$), upon
replacing the index pair $aM$ with $\gamma$ and the index pair $Pq$
with $\kappa$. For
 the most general 12$\times$12 matrix encountered in
this paper, which  has the form
\begin{equation}
\breve{\*A} = \left(%
\begin{array}{cccccccccccc}
  A_{1,1} & A_{1,2} & A_{1,3} & 0 & 0 & 0 & 0 & 0 & 0 & 0 & 0 & A_{1,12} \\
  A_{2,1} & A_{2,2} & A_{2,3} & 0 & 0 & 0 & 0 & 0 & 0 & 0 & 0 & A_{2,12} \\
  A_{3,1} & A_{3,2} & A_{3,3} & 0 & 0 & 0 & 0 & 0 & 0 & 0 & 0 & A_{3,12} \\
  0 & 0 & 0 & A_{4,4} & 0 & 0 & A_{4,4} & 0 & 0 & 0 & A_{4,11} & 0 \\
  0 & 0 & 0 & 0 & A_{5,5} & 0 & 0 & A_{5,5} & 0 & A_{5,10} & 0 & 0 \\
  0 & 0 & 0 & 0 & 0 & A_{6,6} & 0 & 0 & A_{6,6} & 0 & 0 & 0 \\
  0 & 0 & 0 & A_{4,4} & 0 & 0 & A_{4,4} & 0 & 0 & 0 & A_{4,11} & 0 \\
  0 & 0 & 0 & 0 & A_{5,5} & 0 & 0 & A_{5,5} & 0 & A_{5,10} & 0 & 0 \\
  0 & 0 & 0 & 0 & 0 & A_{6,6} & 0 & 0 & A_{6,6} & 0 & 0 & 0 \\
  0 & 0 & 0 & 0 & A_{10,5} & 0 & 0 & A_{10,5} & 0 & A_{10,10} & 0 & 0 \\
  0 & 0 & 0 & A_{11,4} & 0 & 0 & A_{11,4} & 0 & 0 & 0 & A_{11,11} & 0 \\
  A_{12,1} & A_{12,2} & A_{12,3} & 0 & 0 & 0 & 0 & 0 & 0 & 0 & 0 & A_{12,12} \\
\end{array}%
\right),
\end{equation}
the correspondence between the extended symbol indexes and the
matrix indexes is provided in
 Table \ref{MatrixConv}. The scheme  presented in
 Table \ref{MatrixConv} also relates the extended symbol $\breve{t}_{aM}$ to
  the corresponding column
 12--vector entries $\les \, \breve{\#t} \, \ris_\gamma$.

\begin{table}[h!]
  \centering
\begin{tabular}{|c|c||c|c||c|c||c|c|}
\hline
   $aM$ or $Pq$ & $\gamma$ or $\kappa$ &  $aM$ or $Pq$ & $\gamma$ or $\kappa$& $aM$ or $Pq$& $\gamma$ or $\kappa$ & $aM$ or $Pq$ & $\gamma$ or $\kappa$  \\
  \hline
  \hline
  $11$  & $1$ & $23$ or $32$ & $4$ & $23$ or $32$ & $7$ & $14$ or $41$ &  $10$   \\
  $22$  & $2$ & $13$ or $31$ & $5$ & $13$ or $31$ & $8$ & $24$ or $42$ &  $11$   \\
  $33$  & $3$ & $12$ or $21$ & $6$ & $12$ or $21$ & $9$ & $34$ or $43$ &  $12$   \\
  \hline
\end{tabular}
  \caption{Conversion between extended symbol and matrix notation.}\label{MatrixConv}
\end{table}

We introduce the matrix $\breve{\*A}^{\dagger}$ which plays a role
similar to the matrix inverse insofar as
\begin{equation}
\breve{\*A}^{\dagger}\cdot \breve{\*A}  = \breve{\*A} \cdot
\breve{\*A}^{\dagger} = \mbox{\boldmath$\*\tau$} .
\end{equation}
 Herein,
\begin{equation}
\mbox{\boldmath$\*\tau$} = \le \begin{array}{llll} \*I &
\*0_{\,3\times3} & \*0_{\,3\times3} & \*0_{\,3\times3} \vspace{4pt} \\
\*0_{\,3\times3} & \frac{1}{2} \*I & \frac{1}{2} \*I &
\*0_{\,3\times3} \vspace{4pt} \\
\*0_{\,3\times3} & \frac{1}{2} \*I & \frac{1}{2} \*I &
\*0_{\,3\times3} \vspace{4pt} \\
\*0_{\,3\times3} & \*0_{\,3\times3}  & \*0_{\,3\times3}  & \*I
\end{array}
\ri
\end{equation}
is the 12$\times$12 matrix representation of the extended identity
symbol, with $\*I$ being the 3$\times$3 identity matrix,  and we
have
\begin{equation}
\breve{\*A} \cdot\mbox{\boldmath$\*\tau$} =
\mbox{\boldmath$\*\tau$}\cdot \breve{\*A} = \breve{\*A}.
\end{equation}
The matrix $\breve{\*A}^{\dagger}$ has the form
\begin{equation}
\breve{\*A}^{\dagger} = \left(%
\begin{array}{cccccccccccc}
  \dagger_{1,1} & \dagger_{1,2} & \dagger_{1,3} & 0 & 0 & 0 & 0 & 0 & 0 & 0 & 0 & \dagger_{1,12} \\
  \dagger_{2,1} & \dagger_{2,2} & \dagger_{2,3} & 0 & 0 & 0 & 0 & 0 & 0 & 0 & 0 & \dagger_{2,12} \\
  \dagger_{3,1} & \dagger_{3,2} & \dagger_{3,3} & 0 & 0 & 0 & 0 & 0 & 0 & 0 & 0 & \dagger_{3,12} \\
  0 & 0 & 0 & \frac{\dagger_{4,4}}{2} & 0 & 0 & \frac{\dagger_{4,4}}{2} & 0 & 0 & 0 & \dagger_{4,11} & 0 \\
  0 & 0 & 0 & 0 & \frac{\dagger_{5,5}}{2} & 0 & 0 & \frac{\dagger_{5,5}}{2} & 0 & \dagger_{5,10} & 0 & 0 \\
  0 & 0 & 0 & 0 & 0 & \frac{\dagger_{6,6}}{2} & 0 & 0 & \frac{\dagger_{6,6}}{2} & 0 & 0 & 0 \\
  0 & 0 & 0 & \frac{\dagger_{4,4}}{2} & 0 & 0 & \frac{\dagger_{4,4}}{2} & 0 & 0 & 0 & \dagger_{4,11} & 0 \\
  0 & 0 & 0 & 0 & \frac{\dagger_{5,5}}{2} & 0 & 0 & \frac{\dagger_{5,5}}{2} & 0 & \dagger_{5,10} & 0 & 0 \\
  0 & 0 & 0 & 0 & 0 & \frac{\dagger_{6,6}}{2} & 0 & 0 & \frac{\dagger_{6,6}}{2} & 0 & 0 & 0 \\
  0 & 0 & 0 & 0 & \dagger_{10,5} & 0 & 0 & \dagger_{10,5} & 0 & \dagger_{10,10} & 0 & 0 \\
  0 & 0 & 0 & \dagger_{11,4} & 0 & 0 & \dagger_{11,4} & 0 & 0 & 0 & \dagger_{11,11} & 0 \\
  \dagger_{12,1} & \dagger_{12,2} & \dagger_{12,3} & 0 & 0 & 0 & 0 & 0 & 0 & 0 & 0 & \dagger_{12,12} \\
\end{array}%
\right),
\end{equation}
with entries
\begin{eqnarray}
\dagger_{1,1}&=&(-A_{12,3}A_{2,2}A_{3,12}+ A_{12,2}A_{2,3}A_{3,12}+
A_{12,3}A_{2,12}A_{3,2}- A_{12,12}A_{2,3}A_{3,2}-\nonumber\\
&&A_{12,2}A_{2,12}A_{3,3}+ A_{12,12}A_{2,2}A_{3,3})/\Lambda ,\\
\dagger_{1,2}&=&(A_{1,2}A_{12,3}A_{3,12}- A_{12,2}A_{1,3}A_{3,12}-
A_{1,12}A_{12,3}A_{3,2}+ A_{12,12}A_{1,3}A_{3,2}-\nonumber\\
&&A_{1,2}A_{12,12}A_{3,3}+ A_{1,12}A_{12,2}A_{3,3})/\Lambda ,\\
\dagger_{1,3}&=&(-A_{1,2}A_{12,3}A_{2,12}+ A_{12,2}A_{1,3}A_{2,12}+
A_{1,12}A_{12,3}A_{2,2}- A_{12,12}A_{1,3}A_{2,2}+\nonumber\\
&&A_{1,2}A_{12,12}A_{2,3}- A_{1,12}A_{12,2}A_{2,3})/\Lambda , \\
\dagger_{2,1}&=&(-A_{12,3}A_{2,12}A_{3,1}+ A_{12,12}A_{2,3}A_{3,1}+
A_{12,3}A_{2,1}A_{3,12}- A_{12,1}A_{2,3}A_{3,12}-\nonumber\\
&&A_{12,12}A_{2,1}A_{3,3}+ A_{12,1}A_{2,12}A_{3,3})/\Lambda , \\
\dagger_{2,2}&=&(A_{1,12}A_{12,3}A_{3,1}- A_{12,12}A_{1,3}A_{3,1}-
A_{1,1}A_{12,3}A_{3,12}+ A_{12,1}A_{1,3}A_{3,12}-\nonumber\\
&&A_{1,12}A_{12,1}A_{3,3}+ A_{1,1}A_{12,12}A_{3,3})/\Lambda ,\\
\dagger_{2,3}&=&(-A_{1,12}A_{12,3}A_{2,1}+A_{12,12}A_{1,3}A_{2,1}+
A_{1,1}A_{12,3}A_{2,12}- A_{12,1}A_{1,3}A_{2,12}+\nonumber\\
&&A_{1,12}A_{12,1}A_{2,3}- A_{1,1}A_{12,12}A_{2,3})/\Lambda ,\\
\dagger_{3,1}&=&(A_{12,2}A_{2,12}A_{3,1}- A_{12,12}A_{2,2}A_{3,1}-
A_{12,2}A_{2,1}A_{3,12}+ A_{12,1}A_{2,2}A_{3,12}+\nonumber\\
&&A_{12,12}A_{2,1}A_{3,2}- A_{12,1}A_{2,12}A_{3,2})/\Lambda ,\\
\dagger_{3,2}&=&(A_{1,2}A_{12,12}A_{3,1}- A_{1,12}A_{12,2}A_{3,1}-
A_{1,2}A_{12,1}A_{3,12}+ A_{1,1}A_{12,2}A_{3,12}+\nonumber\\
&&A_{1,12}A_{12,1}A_{3,2}- A_{1,1}A_{12,12}A_{3,2})/\Lambda ,\\
\dagger_{3,3}&=&(-A_{1,2}A_{12,12}A_{2,1}+A_{1,12}A_{12,2}A_{2,1}+
A_{1,2}A_{12,1}A_{2,12}- A_{1,1}A_{12,2}A_{2,12}-\nonumber\\
&&A_{1,12}A_{12,1}A_{2,2}+ A_{1,1}A_{12,12}A_{2,2})/\Lambda ,\\
\dagger_{4,4}&=&\frac{A_{11,11}}{2(A_{11,11}A_{4,4}-
                    A_{4,11}A_{11,4})} , \\
\dagger_{5,5}&=&\frac{A_{10,10}}{2(A_{10,10}A_{5,5}-
                    A_{5,10}A_{10,5})} , \\
\dagger_{6,6}&=&\frac{1}{2A_{6,6}} ,\\
\dagger_{10,10}&=&\frac{A_{5,5}}{(A_{10,10}A_{5,5}-
                     A_{10,5}A_{5,10})} , \\
\dagger_{11,11}&=&\frac{A_{4,4}}{(A_{11,11}A_{4,4}-
                      A_{11,4}A_{4,11})} , \\
\dagger_{12,12}&=&(-A_{1,3}A_{2,2}A_{3,1}+A_{1,2}A_{2,3}A_{3,1}+
A_{1,3}A_{2,1}A_{3,2}- A_{1,1}A_{2,3}A_{3,2}-\nonumber\\
&&A_{1,2}A_{2,1}A_{3,3}+ A_{1,1}A_{2,2}A_{3,3})/\Lambda ,
\end{eqnarray}
\begin{eqnarray}
\dagger_{1,12}&=&(A_{1,3}A_{2,2}A_{3,12}-A_{1,2}A_{2,3}A_{3,12}-
A_{1,3}A_{2,12}A_{3,2}+ A_{1,12}A_{2,3}A_{3,2}+\nonumber\\
&&A_{1,2}A_{2,12}A_{3,3}- A_{1,12}A_{2,2}A_{3,3})/\Lambda , \\
\dagger_{2,12}&=&(A_{1,3}A_{2,12}A_{3,1}- A_{1,12}A_{2,3}A_{3,1}-
A_{1,3}A_{2,1}A_{3,12}+ A_{1,1}A_{2,3}A_{3,12}+\nonumber\\
&&A_{1,12}A_{2,1}A_{3,3}- A_{1,1}A_{2,12}A_{3,3})/\Lambda , \\
\dagger_{3,12}&=&(-A_{1,2}A_{2,12}A_{3,1}+ A_{1,12}A_{2,2}A_{3,1}+
A_{1,2}A_{2,1}A_{3,12}- A_{1,1}A_{2,2}A_{3,12}-\nonumber\\
&&A_{1,12}A_{2,1}A_{3,2}+ A_{1,1}A_{2,12}A_{3,2})/\Lambda , \\
\dagger_{4,11}&=&\frac{A_{4,11}}{2(A_{11,4}A_{4,11}-
                     A_{11,11}A_{4,4})} , \\
\dagger_{5,10}&=&\frac{A_{5,10}}{2(A_{5,10}A_{10,5}-
                   A_{10,10}A_{5,5})}, \\
\dagger_{12,1}&=&(A_{12,3}A_{2,2}A_{3,1}- A_{12,2}A_{2,3}A_{3,1}-
A_{12,3}A_{2,1}A_{3,2}+ A_{12,1}A_{2,3}A_{3,2}+\nonumber\\
&&A_{12,2}A_{2,1}A_{3,3}- A_{12,1}A_{2,2}A_{3,3})/\Lambda , \\
\dagger_{12,2}&=&(-A_{1,2}A_{12,3}A_{3,1}+ A_{12,2}A_{1,3}A_{3,1}+
A_{1,1}A_{12,3}A_{3,2}- A_{12,1}A_{1,3}A_{3,2}+\nonumber\\
&&A_{1,2}A_{12,1}A_{3,3}- A_{1,1}A_{12,2}A_{3,3})/\Lambda , \\
\dagger_{12,3}&=&(A_{1,2}A_{12,3}A_{2,1}-A_{12,2}A_{1,3}A_{2,1}-
A_{1,1}A_{12,3}A_{2,2}+ A_{12,1}A_{1,3}A_{2,2}-\nonumber\\
&&A_{1,2}A_{12,1}A_{2,3}+ A_{1,1}A_{12,2}A_{2,3})/\Lambda , \\
\dagger_{11,4}&=&\frac{A_{11,4}}{2(A_{11,4}A_{4,11}-
                     A_{11,11}A_{4,4})} , \\
\dagger_{10,5}&=&\frac{A_{10,5}}{2(A_{10,5}A_{5,10}-
                     A_{10,10}A_{5,5})} , \\
\end{eqnarray}
where the scalar
\begin{eqnarray}
\Lambda&=&A_{1,12}A_{12,3}A_{2,2}A_{3,1}-
A_{12,12}A_{1,3}A_{2,2}A_{3,1}- A_{1,1}A_{12,3}A_{2,2}A_{3,12}+
A_{12,1}A_{1,3}A_{2,2}A_{3,12}-
\nonumber\\
&&A_{1,12}A_{12,3}A_{2,1}A_{3,2}+ A_{12,12}A_{1,3}A_{2,1}A_{3,2}+
A_{1,1}A_{12,3}A_{2,12}A_{3,2}-
A_{12,1}A_{1,3}A_{2,12}A_{3,2}+\nonumber\\
&&A_{1,12}A_{12,1}A_{2,3}A_{3,2}- A_{1,1}A_{12,12}A_{2,3}A_{3,2}-
A_{1,12}A_{12,1}A_{2,2}A_{3,3}+
A_{1,1}A_{12,12}A_{2,2}A_{3,3}+\nonumber\\
&&A_{12,2}(A_{1,3}A_{2,12}A_{3,1}- A_{1,12}A_{2,3}A_{3,1}-
A_{1,3}A_{2,1}A_{3,12}+ A_{1,1}A_{2,3}A_{3,12}+\nonumber\\
&&A_{1,12}A_{2,1}A_{3,3}- A_{1,1}A_{2,12}A_{3,3})+
A_{1,2}(-A_{12,3}A_{2,12}A_{3,1}+ A_{12,12}A_{2,3}A_{3,1}+\nonumber\\
&&A_{12,3}A_{2,1}A_{3,12}- A_{12,1}A_{2,3}A_{3,12}-
A_{12,12}A_{2,1}A_{3,3}+ A_{12,1}A_{2,12}A_{3,3}).
\end{eqnarray}

\section*{Appendix B}
 The extended  Eshelby symbol appropriate to orthorhombic $mm2$ piezoelectric
materials,   distributed as  ellipsoidal particles with shape
parameters $\lec a, b, c \ric$, is given by
\cite{DunnTaya,Numerical_Eshelby}
\begin{equation}\label{EshelbySint}
S^{(esh)}_{MnAb}=\left\{
\begin{array}{lcr}\displaystyle{
\frac{1}{8\pi}C^{(1)}_{sJAb}\int_{-1}^{+1}d\zeta_3 \int_{0}^{2\pi}
d\omega \; \les \,
F_{mJsn}(\overline{\vartheta})+F_{nJsm}(\overline{\vartheta})\,
\ris},
&& M=m=1,\,2,\,3 \\
&\\\displaystyle{
 \frac{1}{4\pi}C^{(1)}_{sJAb}\int_{-1}^{+1}d\zeta_3
\int_{0}^{2\pi} d\omega \;  \, F_{4Jsn}(\overline{\vartheta})\,}, &
&M=4
\end{array}\right. ,
\end{equation}
wherein
\begin{equation}
\left.
\begin{array}{l}
\displaystyle{
  F_{MJsn}(\overline{\vartheta}) = \overline{\vartheta}_s
  \overline{\vartheta}_n
  K_{MJ}^{-1},\qquad K_{JR}= \overline{\vartheta}_s C^{(1)}_{sJRn} \overline{\vartheta}_n  } \vspace{4pt}\\
\displaystyle{  \overline{\vartheta}_1 = \frac{\zeta_1}{a}, \qquad
 \overline{\vartheta}_2 = \frac{\zeta_2}{b}, \qquad \overline{\vartheta}_3 = \frac{\zeta_3}{c}} \vspace{4pt}
  \\
 \displaystyle{ \zeta_1 = (1-\zeta_3^2)^{1/2}\cos(\omega), \qquad
  \zeta_2=(1-\zeta_3^2)^{1/2}\sin(\omega), \qquad
  \zeta_3=\zeta_3}
  \end{array} \right\}.
\end{equation}
The integrals in eqs. \r{EshelbySint} can be evaluated using
standard numerical methods \c{num_methods}.

The conversion from the extended Eshelby symbol $S^{(esh)}_{MnAb}$
to the extended Eshelby 12$\times$12 matrix, namely $\*S^{(Esh)}$,
follows the scheme described in Appendix~A.

\vspace{15mm} \noindent {\bf Acknowledgement:} AJD is supported by
an \emph{Engineering and Physical Sciences Research Council} (UK)
 studentship.

\newpage

\newpage

\newpage

\newpage

\newpage

\end{document}